\documentclass[english, letterpaper, 11pt]{article}

\usepackage{fourier}
\usepackage[T1]{fontenc}

\usepackage[margin=1in]{geometry}
\usepackage{amssymb}
\usepackage{amsmath}
\usepackage{amsmath} 
\usepackage{mathrsfs}
\usepackage{xfrac}
\usepackage{bbm} 
\usepackage{scrextend} 
\usepackage{natbib}
\usepackage{multirow} 
\usepackage{graphicx} 
\usepackage{color}
\usepackage[table,xcdraw]{xcolor}
\usepackage{booktabs} 
\usepackage[labelsep=period]{caption} 
\usepackage[normalem]{ulem} 
\usepackage[hang,flushmargin]{footmisc}
\usepackage[hidelinks]{hyperref}
\usepackage{rotating} 
\usepackage{rotating} 
\usepackage{scalefnt} 
\usepackage{setspace} 
\usepackage{soul} 
\usepackage{enumitem}
\usepackage[labelformat=simple]{subcaption}

\newtheorem{theorem}{Theorem}

\newtheorem{assumption}[theorem]{Assumption}

\newtheorem{proposition}[theorem]{Proposition}
\newtheorem{remark}[theorem]{Remark}

\allowdisplaybreaks

\usepackage{authblk} 

\numberwithin{equation}{section}
\newsavebox{\overlongequation}
\newenvironment{longeq}
{\begin{displaymath}\begin{lrbox}{\overlongequation}$\displaystyle}
{$\end{lrbox}\makebox[0pt]{\usebox{\overlongequation}}\end{displaymath}}

\begin{document}

\title{\huge A System Approach to Structural Identification of Production Functions with Multi-Dimensional Productivity\thanks{\textbf{Correspondence}: Emir Malikov, Lee Business School, University of Nevada, Las Vegas, Las Vegas, NV 89154-6005. Email: emir.malikov@unlv.edu. }}
\author[1]{\sc \vspace{0.5cm} Emir Malikov}
\author[2]{\sc Shunan Zhao}
\author[3]{\sc Jingfang Zhang}
\affil[1]{\small University of Nevada, Las Vegas} 
\affil[2]{\small Oakland University} 
\affil[3]{\small University of Kentucky} 
	
\date{\small September 1, 2022}
\maketitle
	
\begin{abstract}
\noindent There is growing empirical evidence that firm heterogeneity is technologically non-neutral. This paper extends Gandhi et al.’s (2020) proxy variable framework for structurally identifying production functions to a more general case when latent firm productivity is multi-dimensional, with both factor-neutral and (biased) factor-augmenting components. Unlike alternative methodologies, our model can be identified under weaker data requirements, notably, without relying on the typically unavailable cross-sectional variation in input prices for instrumentation. When markets are perfectly competitive, we achieve point identification by leveraging the information contained in static optimality conditions, effectively adopting a system-of-equations approach. We also show how one can partially identify the non-neutral production technology in the traditional proxy variable framework when firms have market power.

		
		

\end{abstract} 
	
\onehalfspacing
\thispagestyle{empty} \addtocounter{page}{0}
\clearpage
	
	
\section{Introduction}
\label{sec:introduction}

Production function and productivity (growth) are fundamental economic concepts the importance of which requires no justification among economists. Their identification however remains a challenge. At the micro level, identifying production functions\textemdash and firm productivity, by extension\textemdash from observational data is not a trivial matter due to the endogeneity issue arising from the fact that firm ``productivity'' capturing such factors like tacit knowledge and managerial quality is unobserved yet must be controlled for because it is correlated with input usage. The literature focused on addressing these methodological issues is vast and uses myriad different approaches, but most consider the case of scalar productivity. Until very recently \citep[most notably][]{dj2018}, few studies have considered the identification of production functions when latent firm productivity is multi-dimensional in that the productivity change may not affect marginal products of all inputs in the same proportion (i.e., neutrally) despite the strong evidence thereof in the data \citep[e.g.,][]{raval2020}. In this paper we develop a structural framework for the proxy variable estimation of production functions with multi-dimensional productivity, including factor-augmenting and -neutral components, that does not rely on the typically unavailable cross-sectional variation in prices for instrumentation. 

Among many methods to tackling endogeneity in the production function context, the proxy variable approach by \citet{op1996} and \citet{lp2003} has become one of the most prevalent estimators in applied productivity research in economics as evidenced, e.g., by over 15,000 Google Scholar citations (as of the time of writing) amassed by these two papers alone. A proxy variable methodology for the structural identification of production functions arguably owes its popularity to not only  good empirical performance but also the relative simplicity of implementation. The conventional proxy variable estimators as well as their many refinements and extensions \citep[e.g.,][]{wooldridge2009,deloecker2013,acf2015,kimetal2019,gnr2013,flynnetal2019,malikovlien,malikovzhao2021} assume that latent firm productivity is scalar and factor-neutral. Then, under some structural timing assumptions, this latent productivity can be expressed as a function of (observable) either the physical investment or intermediate inputs along with other state variables by inverting the corresponding input demand/investment function. This identification strategy relies crucially on the scalar unobservable assumption, whereby there exists a single latent productivity term, which is necessary to ensure the invertibility of demand functions to construct a proxy.

However, the usual assumption of a scalar Hicks-neutral productivity in the production function to capture technological/productivity changes remains inconsistent with many economic theories as well as may be too restrictive in many empirical applications. For example, as \citet{dj2018} point out, the traditional theories of both exogenous and endogenous economic growth rest on the assumption that technological change is non-neutral and, in particular, labor-saving. Large cross-firm heterogeneity in variable input ratios documented in the data is at odds with the factor-neutral productivity too, and the ``biased'' technological change has been widely used to explain changes in the labor share of income \citep[see][]{ zhang2019jde,dj2018,dj2019,oberfieldravel2021}.\footnote{A non-Hicksian technological change is also an important feature of aggregate production functions in some recent macroeconomic studies \citep[e.g., see][]{BaqaeeFarhi2019,BaqaeeFarhi2020}. }
Therefore, the impetus to accommodate non-neutral productivity while estimating production functions is strong and has garnered much attention among economists.

We extend \citeauthor{gnr2013}'s (2020) system-based proxy variable framework for structurally identifying production functions to a more general case when latent firm heterogeneity is multi-dimensional, consisting of factor-neutral and factor-augmenting productivities. Following \citet{dj2018}, to model non-neutral technology we augment the standard proxy variable setup by introducing labor-augmenting (Harrod-neutral) productivity in addition to the Hicks-neutral productivity. Our focus on the labor bias of productivity change is motivated by both its being a key element in growth theory and its inherently unique distinction from other traditional inputs like (physical and human) capital and intermediates, which are all producible. As such, the marginal productivity of inputs other than labor (i.e., capital and materials) is assumed to change equiproportionately at the rate determined by the Hicks-neutral productivity, whereas the (relative) productivity of labor is also affected by biased technology shift. Under separability of variable and dynamic inputs, factor-neutral productivity scales the variable input demands but does not change their ratio, which ensures that the information about Harrod-neutral productivity can be teased out and identified (separately from the Hick-neutral component) from the observed variation in firms' variable input ratios. 

To disentangle the two components of firm productivity (neutral and labor-biased) and to identify the production function, we trade the fully nonparametric formulation in \citet{gnr2013} in favor of a parametric specification of the production technology. In doing so, we are able to explicitly utilize a known functional form of the static first-order conditions for freely varying inputs which enables us to derive a proxy for the labor-augmenting productivity in a closed form and use it to effectively concentrate this non-neutral unobservable out. Namely, we assume that the firm's production function takes a flexible log-quadratic translog specification. We then develop a three-step system-based estimation procedure that makes explicit use of static first-order conditions for flexible inputs and the Markovian properties of both productivity components for structural identification. 

Our model is closely related to \citet{dj2018,dj2019} and \citet{zhang2019jde} who also use the information contained in the mix of flexible inputs to separately identify Hicks- and Harrod-neutral productivities.\footnote{Our paper is also related to \citet{demirer2020} who also considers the problem of identifying production functions with non-neutral multi-dimensional productivity. However, unlike ours, his methodology does not take a structural ``proxy variable'' route but a more atheoretical ``control function'' approach to handling endogeneity-inducing latent firm productivities.} The key and important distinction between their and our methodologies is that we develop an alternative identification scheme that does \textit{not} require external instruments (from outside the production function) such as lagged firm-level variation in (input) prices used in these studies. While it may be suitable for their specific empirical applications, the validity and practicality of using lagged cross-sectional variation in prices for identification is not universal. Not only are such price data typically unavailable or prone to measurement errors in micro-level production datasets \citep{lp2003}, but their use as valid instruments may also be problematic on theoretical grounds \citep[see][]{gm1998,abbp2007,acf2015,malikovlien}. Aside from the concerns about plausible exogeneity of heterogeneous input prices, their strength as instruments also implicitly relies on the strong conditions for the price evolution \citep[see][]{flynnetal2019}. In this paper, we therefore contribute to the literature by developing a methodology that identifies the production function and multi-dimensional productivity even if prices are homogeneous. To achieve identification without external firm-level instruments, we leverage the information contained in static optimality conditions, in effect, adopting a system-of-equations approach. 

Our paper is mainly concerned with the identification of production functions for firms operating in perfectly competitive markets. Although the latter assumption continues to be maintained\textemdash implicitly or explicitly\textemdash by most productivity studies in the literature, which is partly dictated by the lack of  firm-level price data, we also discuss an extension in which we relax this assumption by allowing for monopolistic competition in the output market. In contrast to many other studies that deviate from the perfect competition assumption, our setup does not rely on additional price information or a parameterization of the demand or places restrictions on the production technology in a pursuit of point identification. Instead, we show how one can \textit{partially} identify the production function with non-neutral productivity when firms have market power in the traditional proxy variable framework.

We demonstrate the ability of our estimator to successfully identify multi-dimensional firm productivity through a small set of Monte Carlo simulations and, then, provide an empirical illustration by applying it to the firm-level data from China's leather manufacturing industry. We find that the labor-augmenting productivity behaves quite differently from its factor-neutral counterpart: it shows a larger dispersion and minor growth across years. On the other hand, we also find that foreign direct investment (FDI), which is arguably one of the most important productivity boosters available to firms in developing countries, has both economically and statistically significant effect on labor-saving productivity, whereas its effect size on Hicksian productivity is effectively zero. This suggests that the productivity-enhancing effect of FDI on domestic firms’ productivity has a bias towards labor. At the same time, our estimates provide evidence that productivity change in China's leather industry is, overall, factor-neutral.

The rest of the paper is organized as follows. Section \ref{sec:model} describes the model of production with multi-dimensional firm heterogeneity. We describe our identification strategy in Section \ref{sec:identification}, and the detailed estimation procedure is provided in Section \ref{sec:estimation}. We examine the finite-sample performance of our methodology using simulations and provide an empirical illustration  in Section \ref{sec:finitesample}. The extension to imperfect competition is discussed in Section \ref{sec:extensions}, and Section \ref{sec:conclusion} concludes.


\section{A Model of Firm Production}
\label{sec:model}

This section describes a model of production decisions by a firm in the presence of multi-dimensional productivity. Our model builds upon the conceptual paradigm considered by \citet{dj2018} although we abstract away from their many application-specific nuances with the goal of formulating a generic, versatile framework suitable for application to typical datasets.

Consider the production process of a firm $i$ ($i=1,\dots,n$) in the time period $t$ ($t=1,\dots,T$) in which physical capital $K_{it}$, labor $L_{it}$ and an intermediate input such as materials $M_{it}$ are being transformed into the output $Y_{it}$ via production function $F(\cdot)$ given firm productivity. We differentiate between factor-neutral and factor-biased productivities. Namely, let the firm's production technology take the following form:
\begin{equation}\label{eq:prodfn}
Y_{it} = F(K_{it},\exp\left\{\varphi_{it}\right\}L_{it},M_{it})\exp\left\{\omega_{it}\right\}\exp\left\{\eta_{it}\right\} ,
\end{equation}
where, in addition to the usual log-additive Hicks-neutral productivity $\omega_{it}$, we also allow for Harrod-neutral productivity $\varphi_{it}$ that affects firm output indirectly by ``augmenting'' the labor input. Both can be persistent. The error $\eta_{it}$ is an \textit{ex-post} transitory productivity shock, which is sometimes alternatively interpreted as a classical measurement error in the log-output.

In what follows, we characterize structural assumptions about the firm's technology, productivity, economic environment and its dynamic decision-making process which facilitate a structural identification of the model.

\begin{assumption}\label{assm:1} Among the firm's inputs: {\normalfont (i)} physical capital $K_{it}$ is a dynamic input subject to adjustment frictions; {\normalfont (ii)} labor $L_{it}$ and intermediate inputs $M_{it}$ are freely varying inputs with no dynamic implications.
\end{assumption}

Since physical capital is subject to adjustment costs (e.g., time-to-install), the firm optimizes $K_{it}$ dynamically at time $t-1$ rendering it a predetermined input quasi-fixed at time $t$.  Thus, $K_{it}$ is a state variable with dynamic implications that follows the law of motion: 
\begin{equation}\label{eq:k_lawofmotion}
K_{it}=I_{it-1}+(1-\delta)K_{it-1},
\end{equation}
where $I_{it}$ and $\delta$ are the gross investment and the depreciation rate, respectively. Labor and materials are freely varying and are therefore determined by the firm statically at time $t$, given the already optimized choice of $K_{it}$. This is a fairly standard treatment of inputs in the literature \citep[e.g., see][]{op1996,lp2003,wooldridge2009}. The setup can also be extended to allow for more inputs. The only requirement is that there be at least one freely varying input in addition to labor, which is necessary for identification of labor-augmenting productivity (more on this point below).

\begin{assumption}\label{assm:2}
The production relationship between inputs, Hicks- and Harrod-neutral productivities, and the output takes the form of \eqref{eq:prodfn}. The production function $F(\cdot)$ is {\normalfont (i)} continuous and satisfies the standard neoclassical assumptions, including differentiability, positive monotonicity and concavity in inputs, and {\normalfont (ii)} strongly separable in the partition $(K_{it},(\exp\left\{\varphi_{it}\right\}L_{it},M_{it}))$ as follows:
\begin{equation}\label{eq:prodfn_g}
F(K_{it},\exp\left\{\varphi_{it}\right\}L_{it},M_{it})=G\left(K_{it},H\left(\exp\left\{\varphi_{it}\right\}L_{it},M_{it}\right)\right) ,
\end{equation}
where $H(\cdot)$ is homogeneous of arbitrary degree, and {\normalfont (iii)} of the known parametric functional form.
\end{assumption}

The assumption is that the dynamic inputs are separable from the remaining production-function arguments. Separability is an identifying restriction that ensures information about Harrod-neutral productivity can be teased out and identified from the variation in the firm's material-to-labor ratio which does not depend on the Hicks-neutral productivity.\footnote{It is because of this reliance on a statically optimized input ratio that we require at least two freely varying inputs.} This restriction on production technology imposes that the marginal rate of technical substitution between the two variable inputs (materials and labor) does not depend on dynamic inputs. Whether explicit or not, the same assumption is made in the majority of productivity studies using proxy variable estimators, in line with popular practice, when assuming that the technology is Cobb-Douglas or Constant Elasticity of Substitution (CES). The latter is also the case for \citet{dj2018} and \citet{zhang2019jde}. 

In line with the literature, we model persistent firm productivities as first-order Markov processes which we endogenize \`a la \citet{dj2013}, \citet{deloecker2013} and \citet{malikovzhao2021} by incorporating productivity-enhancing and/or ``learning'' activities of the firm. To keep our model as general as possible, we denote all such activities via generic variables $X_{it}$ and $Z_{it}$ which, depending on the empirical application of interest, may measure the firm's R\&D expenditures, FDI exposure, export status/intensity, etc. Letting $\Xi_{it}$ be the information set available to the $i$th firm for making the period $t$ decisions, the dynamics of unobservables are summarized as follows. 

\begin{assumption}\label{assm:3} {\normalfont (i)} Both components of persistent firm productivity $\omega_{it}$ and $\varphi_{it}$ evolve according to their respective controlled first-order Markov processes: $\mathcal{P}_{\omega}(\omega_{it}|\Xi_{it-1}) =$ $\mathcal{P}_{\omega}(\omega_{it}|\omega_{it-1},X_{it-1})$ and $\mathcal{P}_{\varphi}(\varphi_{it}|\Xi_{it-1}) =$ $\mathcal{P}_{\varphi}(\varphi_{it}|\varphi_{it-1},Z_{it-1})$, where some or all elements in $X_{it-1}$ and $Z_{it-1}$ may be common. {\normalfont (ii)} The transitory productivity shock $\eta_{it}$ is an i.i.d.~white noise process: $\mathcal{P}_{\eta}(\eta_{it}|\Xi_{it}) =\mathcal{P}_{\eta}(\eta_{it})$.
\end{assumption}

The Markov assumptions imply the following regressions for $\omega_{it}$ and $\varphi_{it}$:
\begin{align}
	\omega_{it}&=r_{\omega}\left(\omega_{it-1},X_{it-1}\right)+\zeta_{\omega,it}, \label{eq:productivity_law_omega} \\
	\varphi_{it}&=r_{\varphi}\left(\varphi_{it-1},Z_{it-1}\right)+\zeta_{\varphi,it}, \label{eq:productivity_law_varphi}
\end{align}
where $r_{\omega}(\cdot)$ and $r_{\varphi}(\cdot)$ are the conditional-mean functions of $\omega_{it}$ and $\varphi_{it}$, respectively; and $\zeta_{it}^{\omega}$ and $\zeta_{\varphi,it}$ are mean-zero unanticipated random innovations:
$ \mathbb{E}\left[\zeta_{\omega,it} | \Xi_{it-1}\right]=\mathbb{E}\left[\zeta_{\omega,it}  | \omega_{it-1},X_{it-1}\right]=\mathbb{E}\left[\zeta_{\omega,it}\right]=0$ and $ \mathbb{E}\left[\zeta_{\varphi,it} | \Xi_{it-1}\right]=\mathbb{E}\left[\zeta_{\varphi,it}  | \varphi_{it-1},Z_{it-1}\right]=\mathbb{E}\left[\zeta_{\varphi,it}\right]=0$. Also note that Assumption \ref{assm:3}(i) places no restriction on the correlation between two productivity components $\omega_{it}$ and $\varphi_{it}$. The two may correlate via productivity-modifying ``controls'' and the innovations that may reasonably be expected to be positively correlated.

The evolution processes in \eqref{eq:productivity_law_omega}--\eqref{eq:productivity_law_varphi} implicitly assume that productivity-enhancing activities and learning affect future firm productivity with a delay, which is why the dependence of productivities on their controls $X_{it}$ and $Z_{it}$ is lagged implying that the improvements in firm productivity take a period to materialize. In $\mathbb{E}\left[\zeta_{\omega,it} | \Xi_{it-1}\right]=\mathbb{E}\left[\zeta_{\varphi,it} | \Xi_{it-1}\right]=0$, we effectively assume that firms do not adjust their productivity-modifying activities in light of expected \textit{future} innovations in their productivity, which rules out their ability to systematically predict future shocks. Since random innovations $\zeta_{\omega,it}$ and $\zeta_{\varphi,it}$ represent uncertainty in productivity evolution as well as uncertainty in the success of productivity-modifying activities, the firm relies on its knowledge of \textit{contemporaneous} productivities $\omega_{it-1}$ and $\varphi_{it-1}$ when choosing the optimal level of $X_{it-1}$ and $Z_{it-1}$ at time $t-1$ while being unable to anticipate next period's productivity innovations. Those innovations ($\zeta_{\omega,it}$ and $\zeta_{\varphi,it}$) are realized after both $X_{it-1}$ and $Z_{it-1}$ have already been chosen. Analogous timing assumptions are commonly made in the production-function models with controlled productivity processes \citep{vanbiesebroeck2005,dj2013,deloecker2013,malikovetal2020,malikov2021jom}: they render the firm's \textit{past} productivity-modifying activities mean-orthogonal to random innovations at time $t$, thereby helping the identification of the learning effects.

\begin{assumption}\label{assm:4}
Risk-neutral firms maximize the discounted stream of life-time profits in perfectly competitive output and factor markets with homogeneous prices.
\end{assumption}

Following the bulk of the literature, we assume perfectly competitive markets implying that firms are price-takers which, in theory, rules out any operationable firm-level variation in prices. In what follows, we therefore omit prices from the list of relevant determinants entering the firm's decision equations: they are implicitly represented by the firm-common time index. As noted earlier, we aim to develop an estimator that does \textit{not} require firm-level price information typically unavailable in most firm- or plant-level production datasets. Having said that, we discuss ways to relax this assumption and allow for monopolistic power in the output market in Section \ref{sec:extensions}.

With this structural setup, the firm's dynamic optimization problem is described by
\begin{align}\label{eq:bellman}
\mathbb{V}_{t}\big(K_{it},\omega_{it},\varphi_{it}\big) = \max_{I_{it},X_{it},Z_{it}} \Big\{ &\ \Pi_{t}(K_{it},\omega_{it},\varphi_{it}) -
\text{C}^{I}_{t}(I_{it},K_{it}) -
\text{C}^{X}_{t}(X_{it},X_{it-1})-
\text{C}^{Z}_{t}(Z_{it},Z_{it-1})\ + \notag  \\
&\ \beta\mathbb{E}\Big[\mathbb{V}_{t+1}\big(K_{it+1},\omega_{it+1},\varphi_{it+1}\big) \Big| \Xi_{it},I_{it},X_{it},Z_{it}\Big]\, \Big\} ,
\end{align}
where $\beta$ is a time discount factor; $(K_{it},\omega_{it},\varphi_{it})'\in \Xi_{it}$ are the state variables; $\Pi_{t}(\cdot)$ is the value function corresponding to the static profit-maximization problem in \eqref{eq:profitmax}, i.e., a ``short-run'' restricted profit function; and $\text{C}^{\kappa}_t(\cdot)$ is the cost function for capital ($\kappa=I$) and productivity-enhancing activities ($\kappa=\{X,Z\}$). In the above optimization problem, albeit also with dynamic implications, the levels of productivity-enhancing activities $X_{it+1}$ and $Z_{it+1}$ are chosen by the firm contemporaneously at time $t+1$ unlike the level of $K_{it+1}$ which is a delayed decision made at time $t$ (via $I_{it}$).  This allows for persistence in productivity-enhancing activities but does not force them to be subject to adjustment frictions that would also render them delayed and, hence, predetermined.\footnote{For clarity, if the optimal decision concerning production in period $t$ is affected by its history, then that decision is said to be ``dynamic.'' If, due to adjustment frictions, a decision concerning production in period $t$ is effectively made at $t-1$, then we say it is ``predetermined.'' In this nomenclature, $K_{it}$ is dynamic and predetermined, whereas $X_{it}$ and $Z_{it}$ are dynamic but chosen at time $t$.} This distinction is important because it does not rule out a contemporaneous correlation between firm productivities $(\omega_{it},\varphi_{it})'$ and $(X_{it},Z_{it})'$. Then, solving \eqref{eq:bellman} for $I_{it}$, $X_{it}$ and $Z_{it}$ yields their respective optimal policy functions in terms of the firm's state variables.

Technically, our methodology can be formulated without explicit formalization of the firm's dynamic decisions by only describing its static optimization. We opt to spell out the dynamics, however, to structurally contextualize the predeterminedness of fixed inputs and productivity-modifying activities with respect to the innovations in productivity which, otherwise, would have had to be assumed prima facie.


\section{Identification}
\label{sec:identification}

The estimation of the production function in \eqref{eq:prodfn} is not trivial because of the latent nature of firm productivity. In our case, the problem is further complicated by the fact that unobserved productivity is two-dimensional. Omitting $\omega_{it}$ and $\varphi_{it}$ from the production-function regression is ill-advised because it would lead to an endogeneity problem given that firm productivities are correlated with inputs. We tackle this problem by adopting a control function approach \`a la \citet{op1996} and \citet{lp2003}. Specifically, we consider the identification and estimation of the production function \eqref{eq:prodfn} by building on \citeauthor{gnr2013}'s (2020) methodology, which we generalize to accommodate multi-dimensional firm productivity.

Due to the presence of \textit{multiple} unobservables in \eqref{eq:prodfn} and the fundamentally different manner in which they enhance inputs, to achieve the (separable) identification of $\omega_{it}$ and $\varphi_{it}$ we make use of a parametric-form assumption for the production function. This makes our methodology distinct from \citet{gnr2013} whose approach is fully nonparametric. Forgoing a nonparametric formulation of the production function $F(\cdot)$ in favor of a parametric specification is the price of letting firm productivity not be restricted to a single factor-neutral dimension. We adopt a log-quadratic translog specification for $F(\cdot)$.\footnote{E.g., see \citet{deloeckerwarzynski2012} and \citet{deloeckeretal2016} for recent applications of the translog production functions in the structural proxy estimation.} Namely,
\begin{align}\label{eq:tl}
f(\cdot) =&\ \beta_Kk_{it}+\tfrac{1}{2}\beta_{KK}k_{it}^2+ \beta_Mm_{it}+\tfrac{1}{2}\beta_{MM}m_{it}^2 + \beta_{KM}k_{it}m_{it}\ + \notag \\ 
&\ \beta_L[\varphi_{it}+l_{it}]+\tfrac{1}{2}\beta_{LL}[\varphi_{it}+l_{it}]^2 + \beta_{KL}k_{it}[\varphi_{it}+l_{it}]+\beta_{ML}m_{it}[\varphi_{it}+l_{it}] ,
\end{align}
where the lower-case variables/functions denote the logs of the respective variables/functions.

Under Assumption 2(ii), $\beta_{KM}=\beta_{KL}=0$ and the function needs be normalized to be homogeneous of arbitrary degree in freely varying inputs. Following \citet{dj2019}, we set the degree of homogeneity to $\beta_L+\beta_M$. With this, logging both sides of \eqref{eq:prodfn}, we get the following ``restricted'' translog form:
\begin{align}\label{eq:prodfn_tl}
y_{it} &= \underbrace{\beta_Kk_{it}+\tfrac{1}{2}\beta_{KK}k_{it}^2+  \beta_Mm_{it}+\beta_L[\varphi_{it}+l_{it}] 
-\tfrac{1}{2}\beta_0[m_{it}-\varphi_{it}-l_{it}]^2}_{\overline{y}_{it}}+
\omega_{it}+\eta_{it},
\end{align}
where $\beta_0\equiv-\beta_{MM}=-\beta_{LL}=\beta_{ML}$.

We opt for the translog specification chiefly out of convenience given its linearity in parameters. Other functional forms could also be used; e.g., the nested CES specification preferred by \citet{dj2018} and \citet{zhang2019jde}. We describe how to implement our methodology under this alternative parameterization in Appendix \ref{sec:appx_ces}.


\subsection{A System Approach to Identification}
\label{sec:system_ident}

Since freely varying inputs are non-dynamic, the risk-neutral firm's optimal choice of $L_{it}$ and $M_{it}$ can be modeled statically as the concentrated expected profit-maximization problem subject to the already predetermined optimal choice of the quasi-fixed input $K_{it}$ and both components of persistent firm productivity $\omega_{it}$ and $\varphi_{it}$:
\begin{align}\label{eq:profitmax}
\max_{L_{it},M_{it}}\ P_{t}^Y F(K_{it},\exp\left\{\varphi_{it}\right\}L_{it},M_{it})\exp\left\{\omega_{it}\right\}\theta - P_{t}^L L_{it} - P_{t}^M M_{it},
\end{align}
where $P_{t}^Y$, $P_{t}^L$ and  $P_{t}^M$ are respectively the output, labor and material prices that, given the perfect competition assumption, are common to all firms; and $\theta\equiv\mathbb{E}[\exp\{\eta_{it}\}|\ \Xi_{it}]$. The value function corresponding to \eqref{eq:profitmax} yields $ \Pi_{t}(\cdot)$ entering the firm's Bellman equation \eqref{eq:bellman}. The corresponding first-order conditions yield the firm's conditional demand for $L_{it}$ and $M_{it}$. 

Making use of the functional form in \eqref{eq:prodfn_tl}, the static optimality conditions are
\begin{align}
P_{t}^Y \frac{\exp\left\{\overline{y}_{it}\right\}}{L_{it}}\big(\beta_L+\beta_0[m_{it}-\varphi_{it}-l_{it}] \big)\exp\{\omega_{it}\}\theta &=P_{t}^L, \label{eq:foc_l} \\
P_{t}^Y \frac{\exp\left\{\overline{y}_{it}\right\}}{M_{it}}\big(\beta_M-\beta_0[m_{it}-\varphi_{it}-l_{it}] \big)\exp\{\omega_{it}\}\theta &=P_{t}^M. \label{eq:foc_m}
\end{align}

Taking the ratio of \eqref{eq:foc_l} and \eqref{eq:foc_m},  we obtain the equation for the firm's optimal labor-to-material ratio:
\begin{equation}\label{eq:lmratio}
\frac{L_{it}}{M_{it}} = \frac{P_{t}^M}{P_{t}^L}\times\frac{\beta_L+\beta_0[m_{it}-\varphi_{it}-l_{it}]}{\beta_M-\beta_0[m_{it}-\varphi_{it}-l_{it}]},
\end{equation}
which, expectedly, does not depend on factor-neutral productivity $\omega_{it}$ because the latter enhances both inputs equally thereby leaving their ratio unaffected. The input ratio however is affected by the labor-augmenting productivity, since $\varphi_{it}$ changes the \textit{relative} marginal products of labor and materials.

We can solve \eqref{eq:lmratio} for Harrod-neutral productivity $\varphi_{it}$ to arrive at
\begin{equation}\label{eq:lmratio_varphi}
\varphi_{it} = m_{it}-l_{it} +\frac{\beta_L}{\beta_0}-\left(\frac{\beta_L+\beta_M}{\beta_0}\right) S^L_{it},
\end{equation}
where $S^L_{it}\equiv P_{t}^LL_{it}/\big(P_{t}^LL_{it}+P_{t}^MM_{it}\big)$ is the labor share of the firm's variable input cost. This expression is an operationable proxy for unobservable $\varphi_{it}$.

\textsl{First step}.\textemdash We first identify the sum of $\beta_L$ and $\beta_M$ coefficients as well as nuisance parameter $\theta$ and random productivity shocks $\{\eta_{it}\}$. To do so, we transform static first-order conditions in \eqref{eq:foc_l} and \eqref{eq:foc_m} by taking their logs and subtracting \eqref{eq:prodfn_tl} from each of them to obtain the corresponding share equations in logs, i.e.,
\begin{align}
\ln V_{it}^L &= \ln \left( \theta\beta_0\left[\frac{\beta_L}{\beta_0}+ m_{it}-\varphi_{it}-l_{it}\right] \right) - \eta_{it}, \label{eq:share_l} \\
\ln V_{it}^M &= \ln \left( \theta\beta_0\left[\frac{\beta_M}{\beta_0}- m_{it}+\varphi_{it}+l_{it}\right] \right) - \eta_{it}, \label{eq:share_m}
\end{align}
where $V_{it}^L \equiv P_{t}^LL_{it}/\big(P_{t}^YY_{it}\big)$ and $V_{it}^M \equiv P_{t}^MM_{it}/\big(P_{t}^YY_{it}\big)$ are the nominal shares of labor and material costs in total revenue, respectively. 

To operationalize these equations into the estimating regression equations, we need to tackle the unobservable $\varphi_{it}$ appearing on the right-hand size of \eqref{eq:share_l}--\eqref{eq:share_m}. Failure to control for it would lead to endogeneity due to the correlation with Harrod-neutral productivity and freely varying inputs. We control for $\varphi_{it}$ using the material-to-labor ratio proxy function. That is, substituting for $\varphi_{it}$ in either one of the two log-share equations using the expression in \eqref{eq:lmratio_varphi}, we obtain the following variable-input-cost-to-revenue equation in logs:
\begin{align}\label{eq:fst}
\ln R_{it} &= \ln \big( \theta\left[\beta_L+\beta_M\right] \big) - \eta_{it},
\end{align}
where $R_{it}\equiv \big(P^L_{t}L_{it}+P^M_{t}M_{it}\big)/\big(P^Y_{t}Y_{it}\big)$.
The cost-to-revenue ratio $R_{it}$ is observable in the data, and the construction thereof does not require firm-level price data: the information on total flexible input expenditures and total revenue suffices. 

The variable-input-cost-to-revenue equation in \eqref{eq:fst} is useful in that it enables us to identify $\beta_L+\beta_M$, both elements of which enter the production function of interest in \eqref{eq:prodfn_tl}, using the observable information about expenditures on flexible inputs and revenue. Specifically, we first identify a ``scaled'' sum of these two translog coefficients $\theta\times[\beta_L+\beta_M]$ using the  moment condition $\mathbb{E}[\eta_{it} | \Xi_{it}] =0$, from which we have that
\begin{equation}\label{eq:fst_ident_theta}
\ln \big( \theta\left[\beta_L+\beta_M\right] \big) = \mathbb{E}[\ln R_{it}].
\end{equation}

To identify $\beta_L+\beta_M$ net of constant $\theta$, note that $\theta$ can be identified via $
\theta\equiv \mathbb{E}\left[\exp\left\{ \eta_{it}\right\} \right]=\mathbb{E}\left[\exp\left\{ \mathbb{E}[\ln R_{it}]-\ln R_{it}\right\} \right]$, which allows us to isolate $\beta_L+\beta_M$ as follows:
\begin{equation}\label{eq:fst_ident}
\beta_L+\beta_M = \frac{\exp\{\mathbb{E}[\ln R_{it}]\}}{\mathbb{E}\left[\exp\left\{ \mathbb{E}[\ln R_{it}]-\ln R_{it}\right\} \right]}.
\end{equation}
Let the identified $\beta_L+\beta_M$ be denoted as $\delta_{LM}=\beta_L+\beta_M$.

\textsl{Second step}.\textemdash Next, we show how to separate $\beta_L$ and $\beta_M$ and identify $\beta_0$. We utilize the Markov assumption about labor-augmenting productivity. More concretely, with $\delta_{LM}$ already identified in the first step, the proxy function for $\varphi_{it}$ in \eqref{eq:lmratio_varphi} contains only two unknown parameters: $\beta_0$ and $\beta_L$. Substituting this partly identified $\varphi_{it}(\beta_0,\beta_L)$ expression for $\varphi_{it}$ in the Markov productivity evolution process in \eqref{eq:productivity_law_varphi} and treating $\delta_{LM}$ as an observable, we obtain 
\begin{align}\label{eq:sst}
m_{it}-l_{it} +\frac{\beta_L}{\beta_0}-\frac{\delta_{LM}}{\beta_0} S^L_{it}=
r_{\varphi}\left(\left[m_{it-1}-l_{it-1} +\frac{\beta_L}{\beta_0}-\frac{\delta_{LM}}{\beta_0} S^L_{it-1}\right],Z_{it-1}\right)+\zeta_{\varphi,it} ,
\end{align}
which identifies $(\beta_0,\beta_L)'$ as well as the mean productivity function $r_{\varphi}(\cdot)$ on the basis of
\begin{equation}\label{eq:sst_ident}
\mathbb{E}\left[ \zeta_{\varphi,it} |\ 1,m_{it-1}-l_{it-1},S^L_{it-1},Z_{it-1} \right] = 0.
\end{equation}

Note that the nonlinear equation in \eqref{eq:sst} technically contains an endogenous regressor $S^L_{it}$ which is not mean-orthogonal to the innovation $\zeta_{\varphi,it}$ because the former includes information on the choice of both $L_{it}$ and $M_{it}$ which are decided by the firm after $\zeta_{\varphi,it}$ is realized (i.e., after $\varphi_{it}$ is updated). This however does \textit{not} impede identification of \eqref{eq:sst} because $S^L_{it}$ is not a ``free'' regressor but enters the equation subject to a parameter restriction whereby the coefficient thereon is the same as that on weakly exogenous $S^L_{it-1}$. No external instrumentation for $S^L_{it}$ is therefore needed. 

To make our identification arguments more transparent, let unknown function $r_{\varphi}(\cdot)$ be linear and, since it can only be identified up to a constant, normalize $r_{\varphi}(0)=0$.\footnote{Since productivity/efficiency measurements are relative, this is merely a restriction which implies a ``normalized'' zero-mean $\varphi$ and which does not affect the relative rank of firms based on efficiency of their labor. Qualitatively, this normalization is akin to the typical no-intercept restriction for production functions because an additive constant cannot be separated from the Hicksian productivity unless the latter is also normalized to have a zero mean.} More concretely, $r_{\varphi}=\rho_1\big[m_{it-1}-l_{it-1} +\frac{\beta_L}{\beta_0}-\frac{\delta_{LM}}{\beta_0} S^L_{it-1}\big]+\rho_2'Z_{it-1}$. Denoting the vector of exogenous instruments $Q_{it-1}=(1,m_{it-1}-l_{it-1},\allowbreak S^L_{it-1},Z_{it-1}')'$,  consider now the identification of equidimensional parameter vector $\alpha=(\beta_0,\beta_L,\rho_1,\rho_2')'$ in the following nonlinear GMM problem: 
\begin{align}\label{eq:sst_ident_gmm}
\alpha_0 = \arg\min_{\alpha} \mathbb{E}[Q_{it-1}\zeta_{\varphi,it}(\alpha)]'W\mathbb{E}[Q_{it-1}\zeta_{\varphi,it}(\alpha)],
\end{align}
where $W$ is a symmetric positive-definite moment-weighting matrix. To see that \eqref{eq:sst_ident_gmm} identifies all parameters in $\alpha$, consider an information matrix
\begin{align}\label{eq:sst_ident_gmm_info}
\Psi(\alpha) &= \mathbb{E}\left[Q_{it-1}\frac{\partial\zeta_{\varphi,it}(\alpha)}{\partial\alpha'}\right] \notag \\
&= 
\mathbb{E}\left[
\begin{pmatrix}
1 \\ m_{it-1}-l_{it-1} \\ S^L_{it-1} \\ Z_{it-1}
\end{pmatrix}
\begin{pmatrix}
-\frac{\beta_L}{\beta_0^2}(1-\rho_1)+\frac{\delta_{LM}}{\beta_0^2}\left[S^L_{it}-\rho_1S^L_{it-1}\right] \\
\frac{1-\rho_1}{\beta_0} \\
m_{it-1}-l_{it-1} +\frac{\beta_L}{\beta_0}-\frac{\delta_{LM}}{\beta_0} S^L_{it-1} \\
Z_{it-1}
\end{pmatrix}'\right]
\end{align}
and note that it is full-rank (we unpack this expression in Appendix \ref{sec:appx_Psi}). Thus, the information matrix for the GMM problem in \eqref{eq:sst_ident_gmm} when evaluated at the true parameter values $\Psi(\alpha_0)$ has a full column rank. All parameters in $\alpha$ are therefore locally identified \citep[see][]{rothenberg1971}. 

Although, in theory, the four instruments in $Q_{it-1}$ are enough to exactly identify the second-step parameters of interest $\alpha$, there is a potential to improve the finite-sample performance if additional valid instruments are included in the estimation. Obvious candidates are the firm's dynamic quasi-fixed inputs $k_{it},k_{it-1},\dots$ which are weakly exogenous with respect to the time $t$ shocks, including $\zeta_{\varphi,it}$, and relevant for the choice of $(S^{L}_{it},S^{L}_{it-1},Z_{it-1}')'$ through both the static and dynamic optimization decisions. These additional instruments are to act as exclusion restrictions to strengthen the moment condition and help in identification. For instance, \citet{kimetal2019} propose a similar simple strategy to robustify the \citet{acf2015} estimation procedure.

Following identification of $\beta_L$, $\beta_M$ is identified from $\beta_M=\delta_{LM}-\beta_L$ as a by-product. We also achieve the identification of Harrod-neutral productivity via $\varphi_{it}=m_{it}-l_{it} +\beta_L/\beta_0-(\beta_L+\beta_M)/\beta_0\times S^L_{it}$.

\textsl{Third step}.\textemdash With $(\beta_0,\beta_L,\beta_M)'$ identified, we have effectively pinpointed the production function in the dimension of its endogenous freely-varying inputs $L_{it}$ and $M_{it}$ thereby addressing the \citet{gnr2013} critique, whereby the endogenous static inputs are lacking valid internal instruments when directly included as regressors in the proxied production function estimation. This is evident by rewriting \textbf{\eqref{eq:prodfn_tl}} with the substitution for $\omega_{it}$ using its Markov process from \eqref{eq:productivity_law_omega} as follows:
\begin{equation}\label{eq:prodfn_tl_noendo}
y_{it}^* = \beta_Kk_{it}+\tfrac{1}{2}\beta_{KK}k_{it}^2+ r_{\omega}\left(\omega_{it-1},X_{it-1}\right) + \zeta_{\omega,it}+ \eta_{it} ,
\end{equation}
where $y_{it}^*\equiv y_{it} - \beta_Mm_{it}-\beta_L[\varphi_{it}+l_{it}] 
+ \tfrac{1}{2}\beta_0[m_{it}-\varphi_{it}-l_{it}]^2=y_{it} - \beta_Mm_{it}-\beta_L[m_{it} +\beta_L/\beta_0-(\beta_L+\beta_M)/\beta_0\times S^L_{it}] 
+\tfrac{1}{2\beta_0}[(\beta_L+\beta_M) S^L_{it} -\beta_L]^2$ is already identified and, hence, equation \eqref{eq:prodfn_tl_noendo} now contains \textit{no} endogenous regressors that need instrumentation. However, we still need to deal with unobservability of $\omega_{it-1}$.

To identify the rest of the production function, i.e., $(\beta_K,\beta_{KK})'$, we proxy for latent Hicks-neutral $\omega_{it-1}$ in \eqref{eq:prodfn_tl_noendo} by inverting the conditional material demand function implied by the static first-order condition in \eqref{eq:foc_m}:
\begin{align}\label{eq:mdemand_omega}
\omega_{it} =&\ \overbrace{\ln \left[ \frac{P_{t}^M}{P_{t}^Y}\right]-\ln \theta- \ln\big(\beta_M-\beta_0[m_{it}-\varphi_{it}-l_{it}] \big) +(1-  \beta_M)m_{it}-\beta_L[\varphi_{it}+l_{it}] 
+\tfrac{1}{2}\beta_0[m_{it}-\varphi_{it}-l_{it}]^2}^{m^*_{it}}
 \notag \\
&\  -\beta_Kk_{it}-\tfrac{1}{2}\beta_{KK}k_{it}^2 ,
\end{align}
where, given the already identified $(\beta_0,\beta_L,\beta_M,\theta)'$ and $\varphi_{it}$ in the first two steps, $m^{*}_{it}$ is observable. Substituting for $\omega_{it-1}$ using this proxy, from \eqref{eq:prodfn_tl_noendo} we derive 
\begin{equation}\label{eq:tst}
y_{it}^* = \beta_Kk_{it}+\tfrac{1}{2}\beta_{KK}k_{it}^2+ r_{\omega}\left(\big[m^*_{it-1}-\beta_Kk_{it-1}-\tfrac{1}{2}\beta_{KK}k_{it-1}^2\big],X_{it-1}\right) + \zeta_{\omega,it}+ \eta_{it},
\end{equation}
where, under our structural assumptions, all regressors are weakly exogenous because the productivity innovation $\zeta_{\omega,it}$ is realized at time $t$ after the firm had already optimized its dynamic input $K_{it}$ (at time $t-1$) for the period $t$ production and, obviously, after the lagged flexible inputs contained inside $m_{it-1}^*$ were chosen. That is, all right-hand-side variables in \eqref{eq:tst} can self-instrument. This identifies $(\beta_K,\beta_{KK})'$ as well as the mean productivity function $r_{\omega}(\cdot)$ from
\begin{equation}\label{eq:tst_ident}
\mathbb{E}\left[ \zeta_{\omega,it}+ \eta_{it} |\ k_{it},k_{it-1},m_{it-1}^*(m_{it-1},l_{it-1}),X_{it-1} \right] = 0.
\end{equation}

In fact, given the exogeneity of regressors,  $\gamma=(\beta_K,\beta_{KK},r_{\omega}(\cdot))'$ can be identified as a solution to the following nonlinear sieve M-problem:
\begin{align}\label{eq:tst_nls}
\gamma_0 = \arg\min_{\gamma} \mathbb{E}\left[\varrho_{it}(k_{it},k_{it-1},m_{it-1}^*,X_{it-1};\gamma)^2\right],
\end{align}
where $\varrho_{it}(k_{it},k_{it-1},m_{it-1}^*,X_{it-1};\gamma)\equiv\zeta_{\omega,it}+ \eta_{it}$ is the residual function from \eqref{eq:tst}.

\setcounter{theorem}{0}
\begin{remark}\label{remark:1}\normalfont
One can alternatively operationalize the third step using the inverted conditional \textit{labor} demand implied by \eqref{eq:foc_l} to construct a proxy function for $\omega_{it}$ using observable $l_{it}^*$:
\begin{align}\label{eq:laborproxy}
\omega_{it} =&\ \overbrace{\ln \left[ \frac{P_{t}^L}{P_{t}^Y}\right]-\ln\theta- \ln\big(\beta_L+\beta_0[m_{it}-\varphi_{it}-l_{it}] \big) +(1-\beta_L)l_{it} - \beta_Mm_{it}-\beta_L\varphi_{it}
+\tfrac{1}{2}\beta_0[m_{it}-\varphi_{it}-l_{it}]^2}^{l^*_{it}}
\notag \\
&\  -\beta_Kk_{it}-\tfrac{1}{2}\beta_{KK}k_{it}^2 ,
\end{align}
or using a convex combination of the two inverted static input demands. Under the model assumptions, all these proxies are numerically equivalent.
\end{remark}
\setcounter{theorem}{4}

With $(\beta_K,\beta_{KK})'$ identified, we can also recover Hicks-neutral productivity either via the proxy in \eqref{eq:mdemand_omega} or from the production function \eqref{eq:prodfn_tl} as
\begin{align}
\omega_{it} = y_{it} - \beta_Kk_{it}-\tfrac{1}{2}\beta_{KK}k_{it}^2-  \beta_Mm_{it}-\beta_L[\varphi_{it}+l_{it}] 
+\tfrac{1}{2}\beta_0[m_{it}-\varphi_{it}-l_{it}]^2-\eta_{it},
\end{align}
with the translog parameters, Harrod-neutral productivity $\varphi_{it}$ and the transitory shock ${\eta}_{it}$ successfully identified in the three steps.

On a final note, our methodology is robust to the \citet{acf2015} critique that focuses on the inability of structural proxy estimators to separably identify the additive production function and Hicksian productivity proxy. Such an issue normally arises in the wake of perfect functional dependence between freely varying inputs appearing both inside the unknown production function and productivity proxy. Our third-step equation \eqref{eq:tst} does not suffer from such a problem because it contains no endogenous variable input on the right-hand side, the corresponding parameters of which have already been identified from the variable-input-cost-to-revenue equation and Harrodian productivity process in the first two steps.


\subsection{Unidentification of the Standard Proxy Approach}
\label{sec:unidentification_standard}

We now show that, were one to pursue the standard proxy variable approach, the production function \eqref{eq:prodfn_tl} would be unidentified in the absence of external instruments (outside the production function) for freely varying inputs.

Normally, to estimate the production function via proxy variable technique, one makes use of the Markovian nature of unobservables and proxies for them by inverting the firm's optimality conditions. Specifically, along the lines of \citeauthor{dj2018}'s (2018) empirical methodology, the estimating model consists of two \textit{stochastic} equations: (\textit{i}) the production function in \eqref{eq:prodfn_tl} combined with the law of motion for Hicksian productivity in \eqref{eq:productivity_law_omega}:
\begin{align}\label{eq:djmodel1}
y_{it} =&\ \beta_Kk_{it}+\tfrac{1}{2}\beta_{KK}k_{it}^2+  \beta_Mm_{it}+\beta_L[\varphi_{it}+l_{it}] 
	-\tfrac{1}{2}\beta_0[m_{it}-\varphi_{it}-l_{it}]^2\ + \notag \\
	&\ r_{\omega}\left(\omega_{it-1},X_{it-1}\right) + \zeta_{\omega,it}+ \eta_{it}
\end{align}
and (\textit{ii}) the the law of motion for Harrodian productivity in \eqref{eq:productivity_law_varphi}:
\begin{align}\label{eq:djmodel2}
\varphi_{it}=r_{\varphi}\left(\varphi_{it-1},Z_{it-1}\right)+\zeta_{\varphi,it},
\end{align}
in both of which the unobservables $\varphi_{it}$ and $\omega_{it}$ are ``controlled'' for using their \textit{deterministic} proxy expressions in \eqref{eq:lmratio_varphi} and \eqref{eq:mdemand_omega}, respectively. 

The two-equation model in \eqref{eq:djmodel1}--\eqref{eq:djmodel2} suffers from endogeneity originating in the correlation between freely varying inputs $m_{it}$ and $l_{it}$ (and functions thereof) and contemporaneous random productivity innovations. The remaining covariates are predetermined and can self-instrument. To identify the model, one is to use a system of moment restrictions on the two errors $(\zeta_{\omega,it}+ \eta_{it},\zeta_{\varphi,it})'$ \`a la
\begin{align}\label{eq:djmodel3}
\mathbb{E}\left[ A_{it}\begin{pmatrix}
\zeta_{\omega,it}+ \eta_{it} \\ \zeta_{\varphi,it}
\end{pmatrix}\right]=0,
\end{align}
where $ A_{it}$ is a block-diagonal matrix of exogenous instruments and their functions.

Following the bulk of literature, one may be tempted to make use of higher-order\footnote{The first-order lags are already utilized in the estimation inside Markov processes and thus self-instrument.} lagged inputs $(k_{it-2},l_{it-2},m_{it-2},\dots)'$ and productivity-modifying controls $(X_{it-2}',Z_{it-2}',\dots)'$ to instrument for $m_{it}$ and $l_{it}$ given that such lags meet the weak exogeneity requirement for valid instruments under the structural assumptions. However, the identification can only be achieved if these additional lags provide additional \textit{relevant} (exogenous) variation for $M_{it}$ after conditioning on the already included self-instrumenting variables. Intuitively, these additional lags must be relevant to meet the rank condition for identification of the model.

It happens that despite the apparent abundance of internal instruments, model \eqref{eq:djmodel1}--\eqref{eq:djmodel2} remains \textit{un}identified because all such relevant instruments for $m_{it}$ and $l_{it}$ that were initially excluded from production function \eqref{eq:prodfn_tl} are now used to proxy for the unobserved $\varphi_{it-1}$ and $\omega_{it-1}$. This is the key argument in the \citet{gnr2013} critique of the proxy estimators. More formally, consider the endogenous $m_{it}$ input which, according to the conditional material demand implied by the first-order condition in \eqref{eq:foc_m}, is given by the following implicit function:
\begin{align}\label{eq:mident}
0=&\ \ln \left[ P_{t}^Y/P_{t}^M\right]+\ln\theta+ \ln\big(\beta_M-\beta_0\big[m_{it}-l_{it}-r_{\varphi}\left(\varphi_{it-1},Z_{it-1}\right)-\zeta_{\varphi,it}\big] \big) -(1-  \beta_M)m_{it} \ +
\notag \\
&\  \beta_L\big[l_{it}+r_{\varphi}\left(\varphi_{it-1},Z_{it-1}\right)+\zeta_{\varphi,it}\big] 
-\tfrac{1}{2}\beta_0\big[m_{it}-l_{it}-r_{\varphi}\left(\varphi_{it-1},Z_{it-1}\right)-\zeta_{\varphi,it}\big]^2 \ + \notag \\
&\ \beta_Kk_{it}+\tfrac{1}{2}\beta_{KK}k_{it}^2 + r_{\omega}\left(\omega_{it-1},X_{it-1}\right) + \zeta_{\omega,it} ,
\end{align}
where we have substituted for $\varphi_{it}$ and $\omega_{it}$ using their respective laws of motion. An analogous expression exists for $l_{it}$ which, when combined with \eqref{eq:mident}, expectedly shows that both static inputs are a function of  $(K_{it},\varphi_{it-1},\omega_{it-1},X_{it-1}', Z_{it-1}',P_{t}^Y,P_{t}^M,P_{t}^L,\zeta_{\varphi,it},\zeta_{\omega,it})'$. Comparing these determinants of $m_{it}$ and $l_{it}$ with the variables entering the two equations in \eqref{eq:djmodel1}--\eqref{eq:djmodel2}, it is evident that the only sources of variation in $m_{it}$ and $l_{it}$, which have not already been included as a self-instrumenting variable, are the prices $(P_{t}^Y,P_{t}^M,P_{t}^L)'$ and the unobservable innovations $\zeta_{\varphi,it}$ and $\zeta_{\omega,it}$. 

Assuming the price-taking behavior of competitive firms, \textit{aggregate} prices $(P_{t}^Y,P_{t}^M,P_{t}^L)'$ provide very little\textemdash however theoretically valid\textemdash identifying variation in practice (even in long panels) as studied by \citet{gnr2013}. This means that, conditional on the already included predetermined variables (or their proxies), there is practically \textit{no} other relevant exogenous variables from within the production model that may be used to instrument for the endogenous $m_{it}$ and $l_{it}$ because, for them to be relevant in predicting $m_{it}$ and $l_{it}$, they would have to correlate with $\zeta_{\varphi,it}$ and $\zeta_{\omega,it}$, which is the only source of ``free'' variation left in static inputs. The correlation with these productivity innovations would however violate the exogeneity requirement thereby invaliding the instruments. Therefore, both flexible inputs lack excluded relevant internal instruments, and the model in \eqref{eq:djmodel1}--\eqref{eq:djmodel2} is \textit{un}identified. 

Note that, although not explicitly discussed, this unidentification problem is overcome by \citet{dj2018} along the lines of their earlier work in \citet{dj2013} by incorporating external instruments such as demand shifters and, importantly, lagged \textit{firm-level} variation in input prices. However, while suitable for their empirical application, the validity and practicality of using lagged firm-level prices for identification is not universal. Not only are the price data often unavailable or prone to measurement errors \citep{lp2003}, but the use of prices may also be problematic on theoretical grounds \citep[see][]{gm1998,abbp2007,acf2015,flynnetal2019}. 

Specifically, the validity of input prices as exogenous instruments is normally justified by invoking the assumption of perfectly competitive markets. However, if firms were indeed price-takers, in theory, one should not observe the firm-level variation in prices and, without such a variation, prices cannot be used as operational instruments. Even with the aggregate prices varying exogenously across space, such a variation may be insufficient for identification as discussed earlier. If a researcher does observe the variation in prices across all individual firms, the latter variation may be reflecting differences in the firms' market power and/or the quality of inputs/outputs. For instance, if the firm-level variation in input prices reflects differential quality in inputs, then random updates in prices that render lagged prices usable instruments\footnote{That is, not perfectly dependent with contemporaneous prices that enter estimating equations directly.} are likely related to productivity innovations because a more productive firm is to use more productive, higher-quality inputs \citep{flynnetal2019}. Thus, be it due to the market power or quality differentials, the variation in prices will then be endogenous to firms' decisions and hence cannot help the identification \citep[also see][]{gnr2013}. Furthermore, putting the issue of exogeneity aside, \citet{flynnetal2019} raise concerns about the strong conditions on the price evolution processes that must be satisfied for the lagged prices to have any strength as instruments. We therefore pursue an alternative identification strategy which does not require \textit{firm-level} variation in prices. 

To achieve identification without external firm-level instruments, we build upon \citeauthor{gnr2013}'s (2020) ideas by effectively augmenting a system in \eqref{eq:djmodel1}--\eqref{eq:djmodel2} with an additional (simultaneous) equation for the optimal variable-input-cost-to-revenue ratio in \eqref{eq:fst}.


\section{Estimation Procedure}
\label{sec:estimation}

We now describe how to empirically implement our identification strategy outlined in Section \ref{sec:system_ident}. We estimate the unknown production-function parameters and the two components of persistent firm productivity $\varphi_{it}$ and $\omega_{it}$ via a three-stage procedure. If the functional form of productivity conditional-mean functions $r_{\omega}(\cdot)$ and $r_{\varphi}(\cdot)$ in the evolution processes \eqref{eq:productivity_law_omega}--\eqref{eq:productivity_law_varphi} is known, the estimation becomes fully parametric which, among other things, can streamline asymptotic inference. For instance, a go-to choice in applied productivity research involving the proxy-variable estimation of production functions is to assume that firm productivity is a linear AR(1) process \citep[e.g., see][]{zhang2017eer,zhang2019jde,kimetal2019,afkls2020,griecopaint,moetal2021}. To maximize impact among practitioners, in what follows, we also assume that both $r_{\omega}(\cdot)$ and $r_{\varphi}(\cdot)$ are linear (parametric) functions.\footnote{We can also justify this linearity as a sieve \textit{approximation} using linear polynomials. However, in this case the estimation will no longer be ``parametric'' but semiparametric. See Appendix \ref{sec:appx_semi}.} We discuss a semiparametric alternative to our estimator in which the unknown functions are approximated using sieves in Appendix \ref{sec:appx_semi}.

In the first step, we consistently estimate $\beta_L+\beta_M$ via a sample analogue of \eqref{eq:fst_ident}:
\begin{equation}\label{eq:fst_est}
\widehat{\delta}_{LM} \equiv\widehat{\left(\beta_L+\beta_M\right)} =  \frac{\exp\{\frac{1}{nT}\sum_{it}\ln R_{it}\}}
{\frac{1}{nT}\sum_{it}\exp\left\{ \left(\frac{1}{nT}\sum_{it}\ln R_{it}\right)-\ln R_{it}\right\}},
\end{equation}
where the denominator is $\widehat{\theta}=\frac{1}{nT}\sum_{it}\exp\left\{ \left(\frac{1}{nT}\sum_{it}\ln R_{it}\right)-\ln R_{it}\right\}$.

We then proceed to the second-step estimation of the Harrodian productivity process in \eqref{eq:sst}. We parameterize the unknown function $r_{\varphi}(\varphi_{it-1},Z_{it-1})$ using a linear function,\footnote{As noted earlier, we normalize $r_{\varphi}(0)=0$ since $\varphi_{it}$ can be identified up to a constant only. In practice, this implies that $r_{\varphi}(\cdot)$ is parameterized using a linear function with no intercept.} with the unobservable $\varphi_{it-1}$ substituted for by the proxy function from \eqref{eq:lmratio_varphi} and $\delta_{LM}$ replaced with its estimate from the first step. The second-step equation is then estimated via nonlinear GMM along the lines of \eqref{eq:sst_ident_gmm}:
\begin{align}\label{eq:sst_ident_gmm_est}
\widehat{\alpha} = \arg\min_{\alpha} \Big[\tfrac{1}{nT}\sum_{it}\mathbb{Q}_{it-1}\epsilon_{it}(D_{it};\alpha)\Big]'W\Big[\tfrac{1}{nT}\sum_{it}\mathbb{Q}_{it-1}\epsilon_{it}(D_{it};\alpha)\Big],
\end{align}
where $\alpha=(\beta_0,\beta_L,\rho_{\varphi,1},\rho_{\varphi,2})'$, $D_{it}=(m_{it}-l_{it},m_{it-1}-l_{it-1},S^L_{it},S^L_{it-1},Z_{it-1}')'$, and the corresponding residual function is 
\begin{align}
\epsilon_{it}(D_{it};\alpha)=&\ m_{it}-l_{it} +\frac{\beta_L}{\beta_0}-\frac{\widehat{\delta}_{LM}}{\beta_0} S^L_{it}\ - \rho_{\varphi,1}\left[m_{it-1}-l_{it-1} +\frac{\beta_L}{\beta_0}-\frac{\widehat{\delta}_{LM}}{\beta_0} S^L_{it-1}\right]-\rho_{\varphi,2}Z_{it-1}.
\end{align}
Following our earlier arguments, here we expand the instrument vector to include capital as additional instruments:  $\mathbb{Q}_{it-1}=(1,m_{it-1}-l_{it-1},S^L_{it-1},Z_{it-1}',k_{it},k_{it-1},\dots)'$. With $\big(\widehat{\beta}_0,\widehat{\beta}_L\big)'$ in hand, we construct $\widehat{\beta}_M=\widehat{\delta}_{LM}-\widehat{\beta_L}$ as well as the estimator of Harrod-neutral productivity via $\widehat{\varphi}_{it}=m_{it}-l_{it} +\widehat{\beta}_L/\widehat{\beta}_0-\widehat{\delta}_{LM}/\widehat{\beta}_0\times S^L_{it}$.

To estimate the third-step equation in \eqref{eq:tst}, we first construct estimators of $y_{it}^*$ and $m_{it}^*$ using the results from steps one and two: $\widehat{y}_{it}^*= y_{it} - \widehat{\beta}_Mm_{it}-\widehat{\beta}_L[\widehat{\varphi}_{it}+l_{it}] 
+\tfrac{1}{2}\widehat{\beta}_0[m_{it}-\widehat{\varphi}_{it}-l_{it}]^2$ and $\widehat{m}^*_{it}=\ln \left[ P_{t}^M/P_{t}^Y\right]-\ln\widehat{\theta}- \ln\big(\widehat{\beta}_M-\widehat{\beta}_0[m_{it}-\widehat{\varphi}_{it}-l_{it}] \big) +(1-  \widehat{\beta}_M)m_{it}-\widehat{\beta}_L[\widehat{\varphi}_{it}+l_{it}] +\tfrac{1}{2}\widehat{\beta}_0[m_{it}-\widehat{\varphi}_{it}-l_{it}]^2$.
Then, using a linear parameterization for $r_{\omega}(\cdot)$ with $\omega_{it-1}$ replaced by its proxy, we estimate $\gamma=(\beta_K,\beta_{KK},\rho_{\omega,0},\rho_{\omega,1},\rho_{\omega_2})'$ via nonlinear sieve least squares in line with \eqref{eq:tst_nls}:
\begin{align}\label{eq:tst_nls_est}
\widehat{\gamma} = \arg\min_{\gamma} \sum_{it}\Bigg[\widehat{y}_{it}^* - \beta_Kk_{it}-\tfrac{1}{2}\beta_{KK}k_{it}^2- \rho_{\omega,0}-\rho_{\omega,1}\big[\widehat{m}^*_{it-1}-\beta_Kk_{it-1}-\tfrac{1}{2}\beta_{KK}k_{it-1}^2\big]-\rho_{\omega,2}X_{it-1}\Bigg]^2.
\end{align}
Using the obtained $(\widehat{\beta}_K,\widehat{\beta}_{KK})'$ estimates, we then construct Hicks-neutral productivity via $\widehat{\omega}_{it} = y_{it} - \widehat{\beta}_Kk_{it}-\tfrac{1}{2}\widehat{\beta}_{KK}k_{it}^2-  \widehat{\beta}_Mm_{it}-\widehat{\beta}_L[\widehat{\varphi}_{it}+l_{it}] 
+\tfrac{1}{2}\widehat{\beta}_0[m_{it}-\widehat{\varphi}_{it}-l_{it}]^2-\widehat{\eta}_{it} $ using $\widehat{\eta}_{it}= \ln \big( \widehat{\theta}\widehat{\delta}_{LM} \big) - \ln R_{it}$ from step one.

The outlined three-step estimator is consistent and asymptotically normal. This is easy to establish by recasting all three steps in a multiple-equation system GMM framework which, conveniently, also permits the derivation of an asymptotic variance-covariance matrix that accounts for a multi-step nature of the estimator \citep[see][]{newey1984}. Essentially, we can rewrite our sequential estimator as a simultaneous multiple-equation system of moment restrictions where the instrument sets vary across equations.

Specifically, referring to all unknown parameters collectively as $\Lambda=\big(\beta_0,\beta_L,\beta_M,\beta_K,\beta_{KK},\theta,\rho_{\varphi,1},\rho_{\varphi,2},$ $\rho_{\omega,0},\rho_{\omega,1},\rho_{\omega,2}\big)'$, the fully expanded third-step error after the substitutions for $y^*_{it}$, $\varphi_{it}$ and $m^*_{it}$ is {\small
\begin{align*}
    r_{it}(\Lambda) =& 
    y_{it} - {\beta}_Mm_{it}-{\beta}_L\left[m_{it} +\frac{\beta_L}{\beta_0}-\frac{\beta_L+\beta_M}{\beta_0} S^L_{it}\right] 
+\frac{1}{2\beta_0}\left[-{\beta_L}+(\beta_L+\beta_M) S^L_{it}\right]^2 
-\beta_Kk_{it}-\tfrac{1}{2}\beta_{KK}k_{it}^2-\rho_{\omega,0}-\\
&  \rho_{\omega,1}\Bigg\{\ln \left[ \frac{P_{t}^M}{P_{t}^Y}\right]- \ln\left[\theta(\beta_L+\beta_M)(1-S^L_{it}) \right] +(1-  {\beta}_M)m_{it}-{\beta}_L\left[m_{it} +\frac{\beta_L}{\beta_0}-\frac{\beta_L+\beta_M}{\beta_0} S^L_{it}\right] +\frac{1}{2\beta_0}\left[-{\beta_L}+(\beta_L+\beta_M) S^L_{it}\right]^2- \\
& \beta_Kk_{it-1}-\tfrac{1}{2}\beta_{KK}k_{it-1}^2\Bigg\}-\rho_{\omega,2}X_{it-1},
\end{align*}}
and we can rewrite the three estimation steps in the form of their equivalent multiple-equation moment restrictions:
\begin{equation}\label{eq:msys}
	\mathbb{E}\left[\boldsymbol{f}(\Lambda)\right] \equiv\mathbb{E} 
	\begin{bmatrix}
		\ln R_{it} - \ln \big( \theta\left[\beta_L+\beta_M\right] \big)  \\
		\exp\left\{ \ln\big( \theta\left[\beta_L+\beta_M\right] \big) - \ln R_{it} \right\} - \theta  \\
		\left(m_{it}-l_{it} +\frac{\beta_L}{\beta_0}-\frac{\beta_L+\beta_M}{\beta_0} S^L_{it}\ - \rho_{\varphi,1}\left[m_{it-1}-l_{it-1} +\frac{\beta_L}{\beta_0}-\frac{\beta_L+\beta_M}{\beta_0} S^L_{it-1}\right]-\rho_{\varphi,2}Z_{it-1} \right)\mathbb{Q}_{it-1} \\
		r_{it}(\Lambda)\dfrac{\partial r_{it}(\Lambda)}{\partial (\beta_K,\beta_{KK},\rho_{\omega,0},\rho_{\omega,1},\rho_{\omega_2})'} 
	\end{bmatrix} = \mathbf{0},
\end{equation}
consisting of three blocks, where the first two moments correspond to the sample estimator of $\delta_{LM}=\beta_L+\beta_M$ and $\theta$ (first block), the middle $\dim(\mathbb{Q})$ moments correspond to the GMM estimation of $\alpha=(\beta_0,\beta_L,\rho_{\varphi,1},\rho_{\varphi,2})'$ in \eqref{eq:sst_ident_gmm_est} (second block) and 
and the remaining 5 orthogonality conditions correspond to the nonlinear least-squares estimation of $\gamma=(\beta_K,\beta_{KK},\rho_{\omega,0},\rho_{\omega,1},\rho_{\omega_2})'$ in \eqref{eq:tst_nls_est} (third block). 

The benefit of interpreting our sequential multi-step estimator as solving a GMM problem corresponding to a system of nonlinear moment equations in \eqref{eq:msys} simultaneously is that the standard large-$n$ limit results for a class of such GMM estimators apply here. Furthermore, using the moment equivalents in \eqref{eq:msys} also facilitates asymptotic inference based on the optimal covariance matrix of GMM estimators $\mathbb{V}\big[\widehat{{\Lambda}}\big] = \big[\mathbb{E}\frac{\partial \boldsymbol{f}({\Lambda})}{\partial {\Lambda}' }\big]^{-1} \mathbb{E} [\boldsymbol{f}({\Lambda})\boldsymbol{f}({\Lambda})' ]
\big[\mathbb{E}\frac{\partial \boldsymbol{f}({\Lambda})}{\partial {\Lambda} }\big]^{-1}$ which, if desired, can also be robustified using usual off-the-shelf methods. Having said that, should one find evaluating the analytical covariance matrix tedious or in the case when asymptotic inference is difficult to justify, bootstrap provides an alternative avenue for hypothesis testing. We discuss how to approximate the sampling distribution of our estimator via wild residual block bootstrap\footnote{Other bootstrap procedures can also be used.} that takes into account a sequential nature of our methodology in Appendix \ref{sec:appx_boot}.

\section{Finite-Sample Performance}
\label{sec:finitesample}

In this section, we examine the finite-sample performance of our methodology. We first demonstrate its ability to successfully identify the multi-dimensional firm productivity and the production-function parameters in a small Monte Carlo
study. Then, we apply our estimator to the firm-level data to provide an empirical illustration in practice.


\subsection{Simulations}
\label{sec:simulations}

We conduct simulations to evaluate the performance of our proposed estimator in finite samples. Our data generating process (DGP) draws from those used by \citet{griecoetal2016}, \citet{gnr2013} and \citet{malikovzhao2021}. More specifically, we consider a balanced panel of $n = \{100, 200, 400,1600\}$ firms operating during $T=10$ time periods.\footnote{We have also experimented with 5 and 50 periods. The results are qualitatively unchanged.} Each panel is simulated 1,000 times. We let that the true production technology take a (restricted) translog form with a two-dimensional firm productivity given in \eqref{eq:prodfn_tl}, where we set $\beta_{K}=0.2$, $\beta_{KK}=-0.01$, $\beta_{M}=0.5$, $\beta_{L}=0.25$ and $\beta_{0}=-0.05$. Given the DGPs for the production variables below, this choice of parameter values facilitates that the monotonicity and curvature properties of the production function are satisfied in the generated data; firms exhibit decreasing returns to scale.

The persistent firm productivity components are generated as follows. To simplify matters, we abstract away from productivity modifiers ($X$ and $Z$) and model the Hicks- and Harrod-neutral productivities as exogenous linear AR(1) processes:
\begin{align}
\omega_{it}&=\rho_{\omega,0}+\rho_{\omega,1}\omega_{it-1}+\zeta_{\omega,it}, \\
\varphi_{it}&=\rho_{\varphi,1}\varphi_{it-1}+\zeta_{\varphi,it},
\end{align}
where we set $\rho_{\omega,0}=0.2$, $\rho_{\omega,1}=0.6$
and $\rho_{\varphi,1}=0.9$. The innovations are drawn independently as 
$\zeta_{\omega,it}\sim\ i.i.d.\ \mathbb{N}(0,\sigma_{{\omega}}^{2})$ and $\zeta_{\varphi,it}\sim\  i.i.d.\ \mathbb{N}(0,\sigma_{{\varphi}}^{2})$, 
with $\sigma_{{\omega}}=\sigma_{{\varphi}}=0.04$. The initial levels of Hicks- and Harrod-neutral productivities $\omega_{i1}$ and $\varphi_{i1}$ are drawn from $\mathbb{U}\left(-1,1\right)$ identically and independently distributed over $i$. The random transitory productivity shocks $\{\eta_{it}\}$ entering the production function are drawn from $\eta_{it}\sim\ i.i.d.\ \mathbb{N}\big(0,\sigma_{\eta}^{2}\big)$ with $\sigma_{\eta}=0.07$.

We assume the following about evolution of the firm's state variables. Physical capital, a dynamic predetermined input, is set to evolve according to $K_{it}=I_{it-1}+\left(1-\delta_{i}\right)K_{it-1}$, where the firm-specific depreciation rates $\delta_{i}\in\left\{ 0.05,0.075,0.10,0.125,0.15\right\}$ are distributed uniformly across $i$, and the investment function takes the following form:
\begin{equation}
I_{it-1}=K_{it-1}^{\iota_{1}}\left[\exp\left\{ \omega_{it-1}\right\} \right]^{\iota_{2}}\left[\exp\left\{ \varphi_{it-1}\right\} \right]^{\iota_{3}},
\end{equation}
where $\iota_{1}=0.8$ and $\iota_{2}=\iota_{3}=0.1$. The initial level of capital is generated as $K_{i1}\sim\ i.i.d.\ \mathbb{U}\left(10,200\right)$.

The optimal labor and materials series (the freely varying inputs) are generated by numerically solving the firm's static first-order conditions in \eqref{eq:foc_l}--\eqref{eq:foc_m} after having
already generated the series of $\left(K_{it},\omega_{it},\varphi_{it}\right)'$ for each firm and time period. When doing so, we normalize $P_{t}^{L}=P_{t}^{M}=\theta\ \forall\ t$ and also intentionally assume away any temporal variation in output prices: $P_{t}^{Y}=1$ for all $t$. In such a scenario,  changes in the firm’s labor-to-materials ratio are driven by the improvements in labor-augmenting productivity only. 

\begin{table}[t]
	\caption{Simulation Results for the Proposed Three-Step Estimator}\label{tab:sims}
	\centering\footnotesize
	\makebox[\linewidth]{
	\begin{tabular}{l rrr r rrr r rrr r rrr}
		\toprule[1pt]
		& \multicolumn{3}{c}{$n=100$} && \multicolumn{3}{c}{$n=200$} && \multicolumn{3}{c}{$n=400$} && \multicolumn{3}{c}{$n=1600$}\\
		& Mean & RMSE & MAE && Mean & RMSE & MAE && Mean & RMSE & MAE && Mean & RMSE & MAE \\
		\cline{2-4} \cline{6-8} \cline{10-12} \cline{14-16}
		$\beta_{K}$	& 0.1900 & 0.5316 & 0.4115 && 0.1705 & 0.3619 & 0.2821 && 0.2077 & 0.2549 & 0.2039 && 
		0.2003 & 0.1277 & 0.1003 \\
		$\beta_{KK}(\times 10)$	& --0.0856 & 0.8633 & 0.6704 && --0.0518 & 0.5927  & 0.4635 && --0.1123 & 0.4167 & 0.3335 && 
		--0.1006 & 0.2096 & 0.1644\\
		$\beta_L$ 		& 0.2502 & 0.0038 & 0.0030 && 0.2502 & 0.0024 & 0.0019 && 0.2500 & 0.0018 & 0.0014  &&
		0.2500 & 0.0009 & 0.0007 \\ 
		$\beta_M$		& 0.4997 & 0.0038 & 0.0030 && 0.4998 & 0.0024 & 0.0019 && 0.4999 & 0.0018 & 0.0014  &&
		0.4999 & 0.0009 & 0.0007\\ 
		$\beta_0(\times 10)$ & --0.5024 & 0.0165 & 0.0124 && --0.5008 & 0.0104 & 0.0081 && --0.5003 & 0.0070 & 0.0056  &&
		--0.5004 & 0.0036 & 0.0029 \\
		$\rho_{\varphi,1}$  & 0.8990 & 0.0135 & 0.0107 && 0.8995 & 0.0094 & 0.0074 && 0.8998 & 0.0062 & 0.0050 &&
		0.8998 & 0.0031 & 0.0024 \\
		$\rho_{\omega,1}$   & 0.5981 & 0.0381 & 0.0305 && 0.5988 & 0.0268 & 0.0211 && 0.5996 & 0.0198 & 0.0156 &&
		0.5998 & 0.0095 & 0.0076 \\
		$\rho_{\omega,0}$   & 0.2082 & 0.6385 & 0.4987 && 0.2338 & 0.4380 & 0.3418 && 0.1891 & 0.3083 & 0.2475 &&
		0.1993 & 0.1546 & 0.1217 \\
		\midrule
		\multicolumn{16}{p{18.4cm}}{\footnotesize {\sc Notes:} The true parameter values are $\beta_{K}=0.20$, $\beta_{KK}\times10=-0.10$, $\beta_{L}=0.25$, $\beta_{M}=0.50$, $\beta_{0}\times10=-0.50$, $\rho_{\varphi,1}=0.90$, $\rho_{\omega,1}=0.60$ and $\rho_{\omega,0}=0.20$. Throughout, $T=10$. } \\
		\bottomrule[1pt] 
	\end{tabular}
	}
\end{table}

We estimate our model via the three-step algorithm outlined in Section \ref{sec:estimation}.\footnote{Following our discussion, we include $k_{it}$ as an additional instrument in the second-step estimation.} For each simulation repetition, we obtain point estimates of the production-function and productivity-process parameters and then report the mean, the root mean squared error (RMSE) and the mean absolute deviation (MAD) of these point estimates computed over 1,000 simulations. Table \ref{tab:sims} reports these simulation results. The results are encouraging and show that, with a modestly large sample size, our methodology recovers the true parameters fairly well, thereby lending support to the validity of our identification strategy. Of all parameters, those obtained in the third step are the most imprecisely estimated. This is unsurprising given that their (nonlinear) estimation relies on the generated regressors estimated in not one but two previous steps all of which are measured with sampling error. But overall, as expected of consistent estimators, the estimation becomes more stable as $n$ grows.


\subsection{Empirical Illustration}
\label{sec:empirical}

We showcase our methodology by applying it to study the multi-dimensional productivity heterogeneity among Chinese manufacturing firms. We let the Hicks- and Harrod-neutral productivities share the same scalar productivity shifter, i.e., $X_{it-1} = Z_{it-1}$ in \eqref{eq:productivity_law_omega} and \eqref{eq:productivity_law_varphi}. Given the well-documented importance of inbound FDI\textemdash as a vehicle of international technological diffusion\textemdash for productivity advances  among domestic firms in the recipient countries \citep[see][for more discussion and references to the related literature]{malikovzhao2021}, we use a measure of the foreign equity share as a productivity-modifying ``control,'' with an objective to examine the potentially differential factor-neutral and labor-saving effects of FDI on firm productivity. Through their foreign investors, domestic firms gain access to intangible productive ``knowledge" assets from abroad such as new technologies, proprietary know-hows, more efficient and innovative marketing and management practices, established relational networks, reputation, etc., which can help boost their productivity. Whether such knowledge/technology transfers are neutral or biased remains, however, unexamined.

\textsl{Data}.\textemdash Our data come from the Chinese Industrial Enterprises Database survey conducted by China’s National Bureau of Statistics (NBS). We focus on the ``leather, fur, feather and the related products'' industry (SIC 2-digit code 19) because China is the largest leather producing country in the world, representing more than a quarter of the annual global production, and is one of the world’s largest leather exporters and importers. Also, a relatively large share of firms (15.2\%) in this industry are foreign-invested. 

The production variables are standard. The firm’s capital stock ($K_{it}$) is the net fixed assets deflated by the price index of investment into fixed assets. Labor ($L_{it}$) is measured as the total wage bill plus benefits deflated by the GDP deflator. Materials ($M_{it}$) are the total intermediate inputs, including raw materials and other production-related inputs, deflated by the purchasing price index for industrial inputs. The output ($Y_{it}$) is defined as the gross industrial output value deflated by the producer price index. The price indices are obtained from NBS and the World Bank. The four variables are measured in thousands of real RMB. Our sample period runs from 1998 to 2007, and the operational sample is an unbalanced panel of 11,167 firms with a total of 31,287 observations.\footnote{We exclude observations with missing values for production variables as well as a small number of likely erroneous observations with the foreign equity share values outside the unit interval.} 

The summary statistics of data are reported in Table \ref{tab:data summary} in Appendix \ref{sec:appx_data}. For all variables, the mean values are much larger than their medians, consistent with their distributions being right-skewed, and their inter-quartile ranges are wide. The large heterogeneity of variable distributions suggests that a more flexible model of the production process, such as ours, is needed to better characterize the production relationship between inputs and output. Also note that all variable statistics are larger for foreign-invested firms ($Z>0$) compared with their wholly-domestically-owned counterparts ($Z=0$). This difference provides a further motivation to incorporate the information about firms' exposure to foreign investment into the analysis, which we accomplish by conditioning productivity evolution processes on the foreign equity share.

\textsl{Results}.\textemdash We report the estimated input elasticities, returns to scale (RTS) and the productivity-process parameters in Table \ref{tab:prod fn parameter}. Note that, albeit parametrically, we model the production function using a log-quadratic form, which is why the estimated elasticities and RTS are observation-specific. As discussed in Section \ref{sec:estimation}, we assume that both $r_{\omega}(\cdot)$ and $r_{\varphi}(\cdot)$ are linear given the wide popularity of an AR(1) assumption for productivity processes among practitioners. Therefore, the estimated productivity parameters are fixed and global. Lastly, following the discussion of our methodology, our main results are obtained using the inverted conditional material demand to proxy for Hicksian productivity. While in theory the estimator is invariant to the choice of a variable input for the role of $\omega$ proxy, whether the latter is the case in practice or not, effectively, provides an indirect test of our model assumptions. The counterpart of Table \ref{tab:prod fn parameter} containing the results obtained using the labor proxy as well as the average of labor and material proxies are provided in Appendix \ref{sec:appx_emprobust}. The estimates differ little and are qualitatively unchanged. 

\begin{table}[t]
  \centering \small
  \caption{Estimates of the Production Function and Productivity Parameters}
  \label{tab:prod fn parameter}
    \begin{tabular}{ccccc}
\toprule[1pt]
\multicolumn{5}{c}{\bf Panel A: Elasticities and Returns to Scale}                                                                     \\
                                        & Mean     & 1st Qu.  & Median                               & 3rd Qu.  \\ \cline{2-5} 
\multicolumn{1}{l}{Capital elasticity}  & 0.0361   & 0.0213   & 0.0368                               & 0.0507   \\
                                        & (0.0080) & (0.0094) & (0.0081)                             & (0.0099) \\
\multicolumn{1}{l}{Labor elasticity}    & 0.0950   & 0.0458   & 0.0864                               & 0.1287   \\
                                        & (0.0003) & (0.0001) & (0.0002)                             & (0.0004) \\
\multicolumn{1}{l}{Material elasticity} & 0.7377   & 0.7041   & 0.7463                               & 0.7869   \\
                                        & (0.0020) & (0.0019) & (0.0021)                             & (0.0022) \\
\multicolumn{1}{l}{RTS}                 & 0.8688   & 0.8540   & 0.8695                               & 0.8834   \\
                                        & (0.0081) & (0.0095) & (0.0082)                             & (0.0100) \\[4pt]
\midrule
\multicolumn{5}{c}{\bf Panel B: Productivity Parameters}  \\ 
\multicolumn{2}{c}{\sl Labor-Augmenting} & & \multicolumn{2}{c}{\sl Factor-Neutral} \\ \cline{1-2} \cline{4-5}
\multicolumn{1}{l}{$\rho_{\varphi,0}$}& 0.0000 &&\multicolumn{1}{l}{$\rho_{\omega,0}$} & 0.6525   \\
                                        &   ---    &          &                                      & (0.1026) \\
\multicolumn{1}{l}{$\rho_{\varphi,1}$}   & 0.5723   &          & \multicolumn{1}{l}{$\rho_{\omega,1}$} & 0.7562   \\
                                        & (0.0244) &          &                                      & (0.0307)  \\
\multicolumn{1}{l}{$\rho_{\varphi,2}$}   & 0.0726   &          & \multicolumn{1}{l}{$\rho_{\omega,2}$} & 0.0020    \\
                                        & (0.0183) &          &                                      & (0.0084)  \\[4pt] 
\midrule
\multicolumn{5}{p{9cm}}{\footnotesize {\sc Notes:} The productivity processes are parameterized as follows:  $\varphi_{it}=\rho_{\varphi,1}\varphi_{it-1}+\rho_{\varphi,2}Z_{it-1}+\zeta_{\varphi,it}$ with $\rho_{\varphi,0}$ normalized to 0, and $\omega_{it}=\rho_{\omega,0}+\rho_{\omega,1}\omega_{it-1}+\rho_{\omega,2}Z_{it-1}+\zeta_{\omega,it}$. Bootstrap standard errors are in parentheses.} \\
\bottomrule[1pt]
\end{tabular}
  \label{tab:addlabel}%
\end{table}%

Per the results in Panel A of Table \ref{tab:prod fn parameter}, the manufacturers of leather, fur, feather and related products in China show a large material elasticity and relatively small elasticities of capital and labor. This oversized importance of intermediate inputs (materials) compared with the capital and labor inputs in the production process of Chinese manufacturing firms is confirmed by other studies that used the same dataset \citep[see, e.g.,][]{brandtetal2017,zhaoetal2020,malikovzhao2021,malikov2021jom}. The implied estimates of RTS have the mean value of 0.8688 (median is 0.8695) with the rather narrow inter-quartile range of 0.029. Statistically, all firms exhibit decreasing returns to scale, i.e., diseconomies of scale. This inference is based on the RTS point estimate being statistically less than 1 at the 5\% significance level. 

Panel B of Table \ref{tab:prod fn parameter} reports the estimated productivity parameters for factor-netural and labor-augmenting components of firm productivity. As discussed in Section \ref{sec:system_ident}, because function $r_{\varphi}(\cdot)$ is identified only up to a constant, the intercept coefficient for $\varphi_{it}$ is normalized to 0. Comparing the autoregressive coefficients $\rho_{\varphi,1}$ and $\rho_{\omega,1}$, we find that Harrod-neutral productivity is not as persistent over time as is Hicks-neutral productivity. Interestingly, we find that the foreign equity share\textemdash a productivity modifier of interest\textemdash has both economically and statistically significant marginal effect on labor-saving productivity, whereas the effect size on factor-neutral productivity is insignificant and effectively zero.  From this we can conclude that, at least in China's leather industry, the productivity-boosting effect of FDI on domestic firms' productivity has a bias towards labor. Thus, better/new technologies and more efficient business practices that firms ``import'' and learn from abroad through their foreign investors, as commonly argued in the FDI literature, appear to be primarily labor-saving as opposed to boosting marginal productivity of all factors. This is a novel empirical finding. More concretely, our point estimate of $\rho_{\varphi,2}$ implies that a 10 percentage point increase in the firm's foreign equity share boosts its expected future labor-biased productivity by about 0.73\%. While this effect size might at first appear to be too modest, it is imperative to remember that $\rho_{\varphi,2}$ only captures a short-run impact of FDI on labor-augmenting productivity (that is, $\partial\mathbb{E}[\varphi_{it+1}|\Xi_{it}]\big/\partial Z_{it}$) and does not account for dynamic effects over time. Obviously, owing to the persistence of productivity, the cumulative implications of receiving more FDI are expected to be bigger in the long run. In fact, under temporal stationarity of $\varphi_{it}$ we have that the long-run effect of a 10 percentage point increase in the firm's foreign equity share on its labor-augmenting productivity is estimated at $0.73/(1-0.5723)\approx 1.7$\%.

\begin{figure}[t]
    \centering
    \begin{subfigure}[b]{0.49\textwidth}
    \centering
         \includegraphics[width=\textwidth]{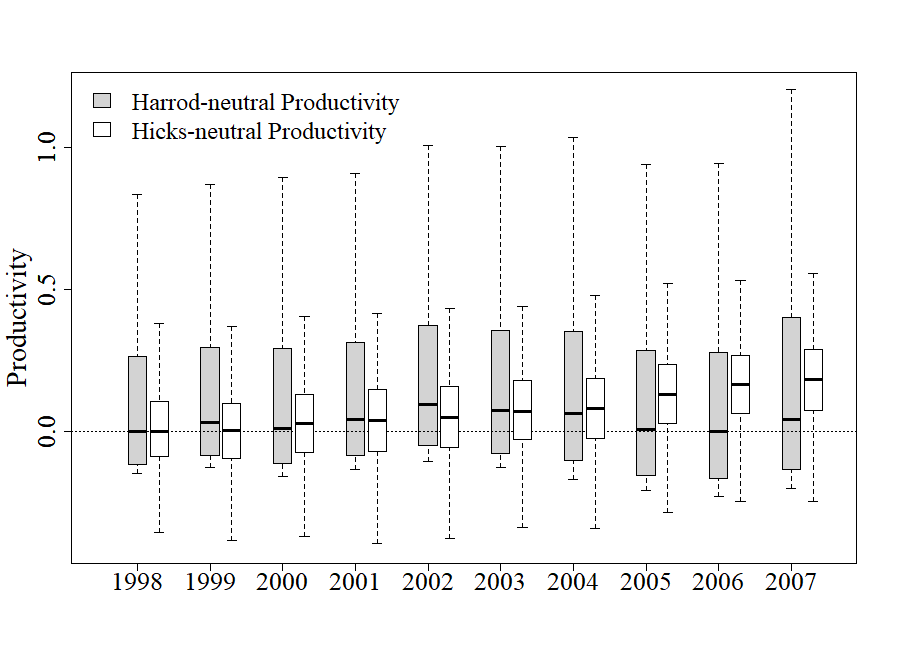}
         \caption{}
         \label{fig:varphi vs omega}
    \end{subfigure}
    \begin{subfigure}[b]{0.49\textwidth}
    \centering
         \includegraphics[width=\textwidth]{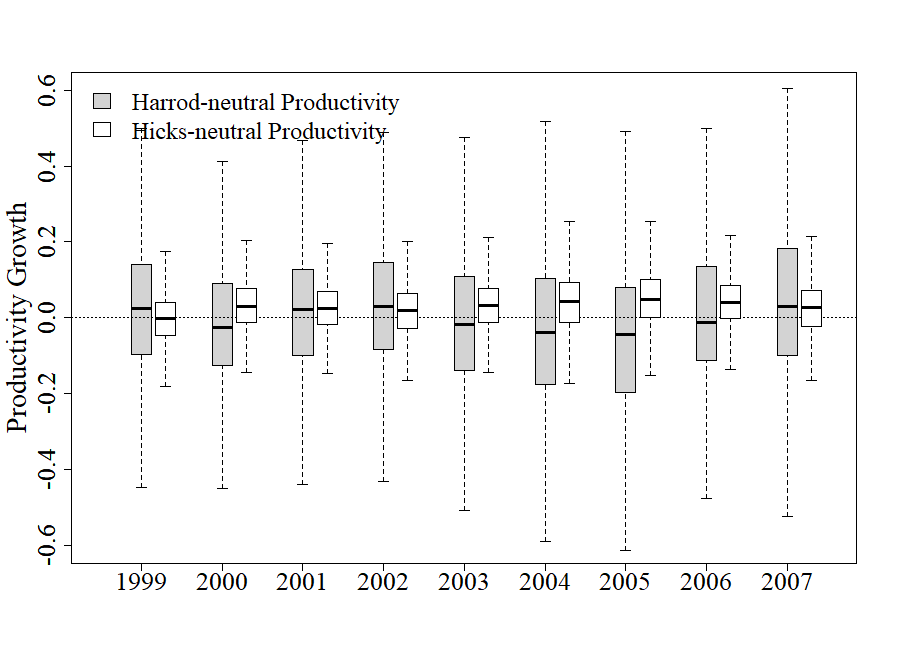}
         \caption{}
         \label{fig:varphi and omega growth}
    \end{subfigure}
    \caption{Distributions of Hicks- and Harrod-Neutral Productivities: (a) in levels, (b) in growth rates}
    \label{fig:varphi_omega}
\end{figure}

Figure \ref{fig:varphi_omega} compares the empirical distributions of the two estimated productivities. We report box-plots of the estimated $\varphi_{it}$ and $\omega_{it}$ by year in Figure \ref{fig:varphi vs omega} and of their annual changes in Figure \ref{fig:varphi and omega growth}. For the ease of comparison, the medians of both productivity terms are normalized to zero in the year 1998. We see that the distributions of Harrod- and Hicks-neutral productivities behave quite differently. First, according to Figure \ref{fig:varphi vs omega}, the labor-augmented productivity has a larger cross-sectional variation, as exhibited by  wider inter-quartile ranges and longer whiskers. Second, in each year, the Hicks-neutral productivity is distributed almost symmetrically across individual firms, whereas the labor-augmented productivity is heavily skewed to the right. Third, based on the medians in each year, the Hicks-neutral productivity steadily shifts up over time, but we cannot say the same about the labor-augmented productivity. Instead, it is mainly the the upper/right whiskers of the labor-augmented productivity that generally shift up over time, suggesting the presence of persistently more labor-efficient firms that keep becoming more productive.  

In Figure \ref{fig:varphi and omega growth}, we plot the box-plots of annual productivity changes in logs, i.e., $\varphi_{it}-\varphi_{it-1}$ and $\omega_{it}-\omega_{it-1}$. Since both the $\varphi_{it}$ and $\omega_{it}$ are attached to the log inputs and output, their changes can be approximately interpreted as the corresponding within-firm productivity growth rates. We see that, compared with factor-neutral productivity, the labor-augmenting productivity also exhibits a larger cross-sectional variation in its growth. The median growth rate of Hicks-neutral productivity is non-negative across all years, whereas that of labor-augmenting productivity oscillates around zero, with essentially a nil cumulative effect. 

\begin{figure}[t]
    \centering
    \begin{subfigure}[b]{0.49\textwidth}
    \centering
         \includegraphics[width=\textwidth]{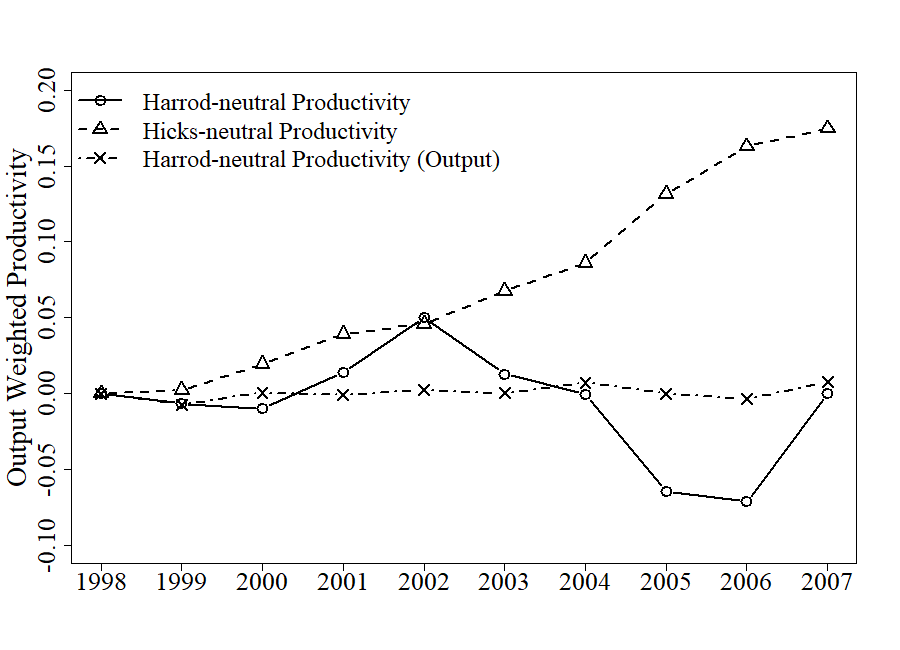}
         \caption{}
         \label{fig:prod trends}
    \end{subfigure}
    \begin{subfigure}[b]{0.49\textwidth}
    \centering
         \includegraphics[width=\textwidth]{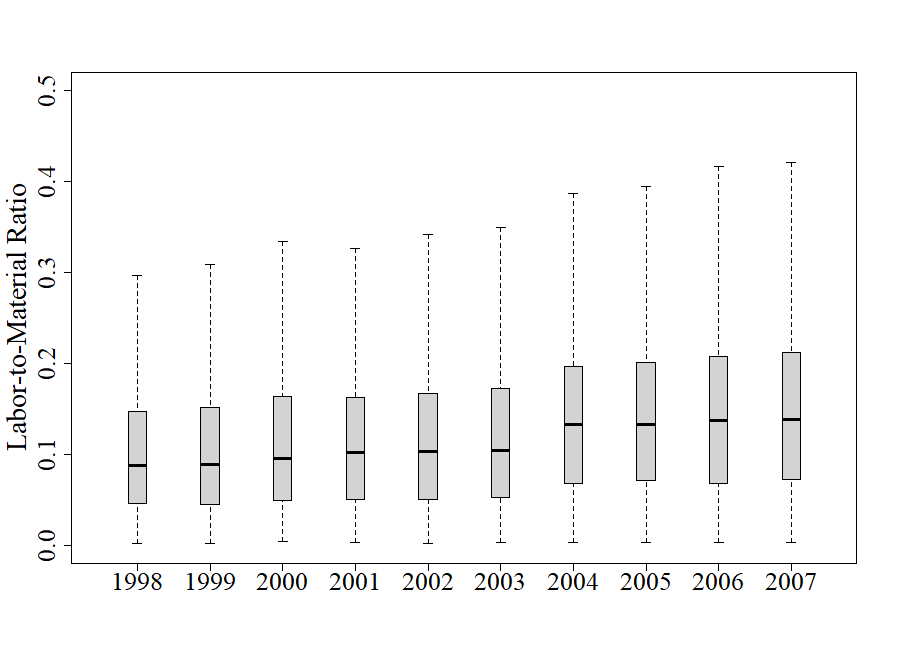}
         \caption{}
         \label{fig:input ratio}
    \end{subfigure}
    \caption{Productivity and Flexible Input Ratio Trends}
    \label{fig:prod_input ratio}
\end{figure}

To see the cumulative and total impact of the growth in these two productivity components on the industry output, we calculate the aggregate industry output-weighted Harrod-neutral and Hicks-neutral productivities and plot their trends in Figure \ref{fig:prod trends}. These two aggregate productivities are depicted using solid-circle and dashed-triangle lines, respectively, and for comparability, both of them are normalized to zero in the year 1998. We find that, during our sample period, the industry-level factor-neutral productivity was steadily rising and increased by around 20\%. However, the aggregate labor-augmenting productivity peaked in 2002 right after China's accession to the WTO and decreased since then, although with a rebound in the last year to a level close to that of the beginning of the sample period. This trend is generally consistent with the labor-to-material ratio box-plot in Figure \ref{fig:input ratio}. Were the labor-augmenting productivity to increase significantly in our sample period, we would have expected the labor-to-material ratio to decrease over time too. Such labor-saving technological advances have been documented by \citet{dj2018} and \citet{zhang2019jde} for different countries/industries. In our case, however, the post-2002 downward trend of labor-augmented productivity is in line with the upward shift in the labor-to-material ratios observed in the data.\footnote{The graph for the output-weighted average labor-to-material ratios looks similar.} The widening whiskers over the years in Figure \ref{fig:input ratio} are also consistent with the increasing variation of labor-augmenting productivity over time documented in Figure \ref{fig:varphi vs omega}.

Now, note that, because the firm's output is log-linear in $\omega_{it}$, the magnitude of Hicks-neutral productivity growth is directly corresponding to the output growth. However, this is not the case for the non-neutral productivity $\varphi_{it}$ that affects output via labor. The marginal effect of $\varphi_{it}$ on the firm's (log) output depends on the labor elasticity. Therefore, to make a fair comparison between Hicks- and Harrod-neutral productivities in terms of their effects on output growth, we follow \citet{dj2018} in computing the product of the labor elasticity $\partial f_{it}/\partial l_{it}$ and $\varphi_{it}$ and refer to it as the labor-augmenting productivity in output terms. In Figure \ref{fig:prod trends}, we plot the industry output-weighted average for it using a dot-dashed-cross line. The line is almost flat, implying that the labor-augmenting productivity had no material effect on the industry output growth during our sample period. Thus, our estimates provide evidence that the overall productivity growth in China's leather industry in 1998--2007 was factor-neutral.


\section{Extension to Imperfect Competition}
\label{sec:extensions}

As argued by \citet{battisti2022skill}, empirical studies have so far produced very limited evidence on the magnitude of non-neutral technological change with market imperfections.\footnote{In their study, \citet{battisti2022skill} present a new approach and estimate the skill-biased technical change from the production side while allowing for labor-market inefficiencies using country-level data.}
Our methodology and its underlying identification scheme presented above are developed under the assumption that firms operate in a perfectly competitive output market. Although this assumption continues to be maintained\textemdash implicitly or explicitly\textemdash by most productivity studies, which is partly dictated by the typical unavailability of firm-level price data,\footnote{This is because, unless the quantity information about output/inputs is observed in the data (which is rare in practice), researchers commonly assume perfect competition with homogeneous prices to justify deflating nominal revenues and expenditures using price indices to obtain real values.}  there has been notable effort recently aimed at extending proxy variable production-function estimators to accommodate market power. Besides usually requiring the data on exogenous demand shifters, most such methods also tend to rely on (observable) exogenous heterogeneous input prices and/or isoelastic demand specifications for identification \citep[e.g., see][]{deloecker2011,deloeckerwarzynski2012,deloeckeretal2016,dj2013,dj2019}. Others resort to restrictions on the production technology such as the constant returns to scale \citep[see][]{flynnetal2019,raval2020} or abandon the structural ``proxy variable'' paradigm in favor of a more ``atheoretical'' control function approach to handling endogeneity-inducing unobservable firm productivity \citep{demirer2020}. 

Motivated by this emerging literature, we show how to relax the perfect competition assumption and allow for monopolistic power in the output market. We do so while retaining the assumption of competitive homogeneous factor prices, given our earlier discussion concerning the use of firm-level variation in input prices for identification. In Appendix  \ref{sec:appx_ext}, we first show \textit{point} non-identification of the production function in \eqref{eq:prodfn_tl} when firms exhibit market power even if one observes exogenous demand shifters in the data. We then discuss how with some additional but quite reasonable assumptions about unobservables one can still \textit{set} identify the production function. We derive this partial identification result without requiring that the information on demand shifters be available and rely on the same set of observables that are used in our methodology for the case of perfectly competitive markets.


\section{Concluding Remarks}
\label{sec:conclusion}

The literature on proxy variable identification of production functions is dominated by models that accommodate a single source of firm heterogeneity: a scalar Hicks-neutral productivity. In this paper, we contribute to the relatively thin but emerging strand of the literature that seeks to generalize these proxy methods to allow for non-neutral production technology by considering the identification of a translog production function when latent firm productivity is multi-dimensional, with biased labor-augmenting and factor-neutral components.
In contrast to the available alternatives, our model can be identified under weaker data requirements, notably, without relying on the firm-level input price information as instruments. This tremendously increases the practical value of our methodology by making it applicable to most production datasets. When markets are perfectly competitive, we achieve point identification by leveraging the information contained in static optimality conditions, in effect, adopting a system-of-equations approach. We also show how one can set identify the production function with non-neutral productivity in the traditional proxy variable framework when firms have market power.


\appendix
\section*{Appendix}
\label{sec:appendix}


\section{Identification under the CES Specification}
\label{sec:appx_ces}

\setcounter{figure}{0} \renewcommand\thetable{\thesection.\arabic{table}}
\setcounter{table}{0} \renewcommand\thefigure{\thesection.\arabic{figure}}

In Section \ref{sec:identification}, we present our identification strategy for a restricted translog production function. To show that the same strategy can also be applied to other production function specifications, here we detail its application to a CES functional form, another widely-used specification for production technology. Specifically, let the firm's production technology $F(\cdot)$ with labor-saving productivity take the following form:
\begin{equation} \label{eq:prod fn. ces}
    F(\cdot) = \left\{ \beta_K K_{it}^{-\frac{1-\sigma}{\sigma}} + \left[\exp\{\varphi_{it}\}L_{it}\right]^{-\frac{1-\sigma}{\sigma}} + \beta_M M_{it}^{-\frac{1-\sigma}{\sigma}} \right\}^{-\frac{\nu \sigma}{1-\sigma}},
\end{equation}
where $\nu$ and $\sigma$ are the elasticities of scale and substitution, respectively; and $\beta_K$ and $\beta_M$ are the distribution parameters, with that corresponding to labor implicitly normalized to unity since it cannot be identified separately from $\varphi_{it}$. 

Among others, the CES form in \eqref{eq:prod fn. ces} has been used by \citet{dj2013}. Besides the usual monotonicity and curvature assumptions, it is easy to confirm that it also satisfies separability per our Assumption  \ref{assm:2}(ii). Namely, $F(\cdot)$ above is a nested CES that is separable in capital as follows: $F(K_{it},\exp\{\varphi\}L_{it},M_{it})= G\left(K_{it},H\left(\exp\{\varphi_{it}\}L_{it},M_{it}\right)\right)$, where
\begin{align*}
	G\left(\cdot\right) &= \left\{\beta_K K_{it}^{-\frac{1-\sigma}{\sigma}}+H\left( \exp\{\varphi_{it}\}L_{it}, M_{it}\right)^{-\frac{1-\sigma}{\sigma}} \right\}^{-\frac{\nu \sigma}{1-\sigma}},  \\
	H\left( \cdot\right) &= \left\{ \left[\exp\{\varphi_{it}\}L_{it}\right]^{-\frac{1-\sigma}{\sigma}} + \beta_M M_{it}^{-\frac{1-\sigma}{\sigma}}  \right\}^{-\frac{ \sigma}{1-\sigma}} ,
\end{align*}
and $H\left( \cdot\right)$ is normalized to be linearly homogeneous (i.e., a unitary scale elasticity within the labor-and-materials pair of inputs), and the elasticity of substitution between capital and the labor-and-materials aggregator is the same as that between labor and materials within the aggregator.

Making use of the ``known'' functional form of $F(\cdot)$, from the static first-order conditions for flexible inputs $L_{it}$ and $M_{it}$ we obtain
the following expression for the Harrod-neutral productivity $\varphi_{it}$:
\begin{equation} \label{eq:lmratio_varphi_ces}
    \varphi_{it} = \frac{1}{1-\sigma} \left( m_{it} - l_{it} \right) - \frac{\sigma}{1-\sigma} \ln(\beta_M) + \frac{\sigma}{1-\sigma} \left( \ln P^M_t - \ln P^L_t \right).
\end{equation}
The above equation is the counterpart of \eqref{eq:lmratio_varphi} when the production function takes a CES functional form, and it provides a proxy for the latent $\varphi_{it}$ expressed as a function of the production function parameters and observed data. 

Next, we combine \eqref{eq:lmratio_varphi_ces} with the law of motion for labor-augmenting productivity, which is specified in \eqref{eq:productivity_law_varphi}, into an estimating equation for parameters of the production function in \eqref{eq:prod fn. ces}. More concretely, substituting $\varphi_{it}$ and $\varphi_{it-1}$ in \eqref{eq:productivity_law_varphi} with \eqref{eq:lmratio_varphi_ces}, we have
\begin{align}\label{eq:sst ces}
&\frac{1}{1-\sigma} \left( m_{it} - l_{it} \right) - \frac{\sigma}{1-\sigma} \ln(\beta_M) + \frac{\sigma}{1-\sigma} \left( \ln P^M_t - \ln P^L_t \right) \notag \\ 
= \quad &r_{\varphi}\left(\left[ \frac{1}{1-\sigma} \left( m_{it-1} - l_{it-1} \right) - \frac{\sigma}{1-\sigma} \ln(\beta_M) + \frac{\sigma}{1-\sigma} \left( \ln P^M_{t-1} - \ln P^L_{t-1} \right) \right],Z_{it-1}\right)+\zeta_{\varphi,it}.
\end{align}

The above equation is the counterpart of \eqref{eq:sst} in the translog case. The two unknown parameters $(\sigma, \beta_M)'$ and the mean function $r_{\varphi}(\cdot)$ can be estimated via nonlinear least squares, given the exogeneity of regressors, viz., $
\mathbb{E}\left[ \zeta_{\varphi,it} \ | \ 1, \ m_{it-1}-l_{it-1}, \ \ln P^M_{t} - \ln P^L_{t}, \ \ln P^M_{t-1} - \ln P^L_{t-1}, \ Z_{it-1} \right] = 0$. Also note that, because of the particular functional form of the CES specification, we are able to recover all parameters pertaining to variable inputs in a single step, whereas in the case of translog, we do so in two steps.

With $\sigma$ and $\beta_M$ identified from \eqref{eq:sst ces}, we can also estimate $\varphi_{it}$ following \eqref{eq:lmratio_varphi_ces}. To see the identification of the remaining parameters in the production function, i.e., $\beta_K$ and $\nu$, take the logarithm of the production function and substitute for $F(\cdot)$ using \eqref{eq:prod fn. ces}: 
\begin{align} \label{eq:prodfn_ces_noendo}
    y_{it} = &-\nu\frac{ \sigma}{1-\sigma} \ln \left\{ \beta_K \overbrace{K_{it}^{-\frac{1-\sigma}{\sigma}}}^{K_{it}^*} + \overbrace{\left[\exp\{\varphi_{it}\}L_{it}\right]^{-\frac{1-\sigma}{\sigma}} + \beta_M M_{it}^{-\frac{1-\sigma}{\sigma}}}^{H_{it}^*} \right\} + \omega_{it} + \eta_{it}, \notag \\
    = & -\nu\frac{\sigma}{1-\sigma} \ln \bigg\{ \beta_K K_{it}^* + H_{it}^* \bigg\} + r_{\omega}\left(\omega_{it-1},X_{it-1}\right) + \zeta_{\omega,it}+ \eta_{it},
\end{align}
where we have replaced $\omega_{it}$ with its law of motion in the second equality. The new variables $K_{it}^*\equiv K_{it}^{-\frac{1-\sigma}{\sigma}}$ and $H_{it}^*\equiv \left[\exp\{\varphi_{it}\}L_{it}\right]^{-\frac{1-\sigma}{\sigma}} + \beta_M M_{it}^{-\frac{1-\sigma}{\sigma}}$ are effectively data because they are defined using observable inputs and the already identified parameters along with labor-augmenting productivity. Also, remember that $\frac{ \sigma}{1-\sigma}$ is known as well.

To address the unobservability of $\omega_{it-1}$, we proxy for it by inverting the conditional material demand function implied by the corresponding static first-order condition for $M_{it}$:\footnote{One can also operationalize this step using the inverted conditional labor demand instead. Both proxies are equivalent.}
\begin{align} \label{eq:mdemand_omega ces}
    \omega_{it} &= \overbrace{\ln P^M_t - \ln P^Y_t + \frac{1}{\sigma} m_{it}  -\ln \left(\beta_M\right)}^{m^*_{it}} + \left( 1+ \nu\frac{ \sigma}{1-\sigma} \right)\ln \bigg\{ \beta_K K_{it}^* + H_{it}^* \bigg\}- \ln \left(\theta \nu \right) 
\end{align}
where $m^*_{it}$ is already identified and, hence, observable. The above proxy for the latent Hicks-neutral productivity is the counterpart of \eqref{eq:mdemand_omega}. Replacing $\omega_{it-1}$ in \eqref{eq:prodfn_ces_noendo} with the lag of \eqref{eq:mdemand_omega ces}, we have the second-step estimating equation that identifies the remaining unknown parameters $(\nu,\beta_K)'$:
\begin{align} 
    y_{it} = & -\nu\frac{ \sigma}{1-\sigma} \ln \left\{ \beta_K K_{it}^{*} + H_{it}^* \right\} + 
    r_{\omega}\left(\left[m^*_{it-1}+ \left( 1+ \nu\frac{ \sigma}{1-\sigma} \right)\ln \bigg\{ \beta_K K_{it-1}^* + H_{it-1}^* \bigg\}- \ln \left(\theta \nu \right) \right],X_{it-1} \right) + \zeta_{\omega,it}+ \eta_{it}. \label{eq:tst ces}
\end{align}

Per the structural assumptions, $K_{it}^*$, $K_{it-1}^*$, $H_{it-1}^*$, $m_{it-1}^*$ and $X_{it-1}$ are all mean-orthogonal to the composite innovation $\zeta_{\omega,it} + \eta_{it}$ in \eqref{eq:tst ces} and can self-instrument. The same cannot be said about $H_{it}^{*}$ that also appears in  the equation, since it includes information on the choice of both $L_{it}$ and $M_{it}$ which are decided by the firm after $\zeta_{\omega,it}$ is realized (i.e., after $\omega_{it}$ is updated). However, analogous to the case with the second-step estimation under the translog specification in \eqref{eq:sst} in Section \ref{sec:system_ident}, the endogeneity of $H_{it}^{*}$ does \textit{not} impede identification of \eqref{eq:tst ces} because $H_{it}^{*}$ is not a ``free'' regressor but enters the equation subject to a parameter restriction whereby its distribution parameter is normalized to 1 and requires no estimation. No external instruments for $H_{it}^{*}$ are therefore needed. As such, we can identify $(\beta_K,\nu)'$ as well as the mean productivity function $r_{\omega}(\cdot)$ from \eqref{eq:tst ces} based on the following moments:
\begin{equation}
\mathbb{E}\left[ \zeta_{\omega,it} + \eta_{it} |\ 1,k_{it}^*,k_{it-1}^*,m_{it-1}^*, H_{it-1}^{*}, X_{it-1} \right] = 0.
\end{equation}


\section{Expanded $\Psi(\alpha)$ from \eqref{eq:sst_ident_gmm_info}}
\label{sec:appx_Psi}

Suppressing the firm index $i$, we have that

{\footnotesize
	\begin{longeq}
		\Psi(\alpha) = 
		\begin{pmatrix}
			\left\{-\frac{\beta_L}{\beta_0^2}(1-\rho_1)+\frac{\delta_{LM}}{\beta_0^2}\left[\mathbb{E}[S^L_{t}]-\rho_1\mathbb{E}[S^L_{t-1}]\right]\right\} &
			\frac{1-\rho_1}{\beta_0} &
			\left\{\mathbb{E}[\breve{m}_{t-1}] +\frac{\beta_L}{\beta_0}-\frac{\delta_{LM}}{\beta_0} \mathbb{E}[S^L_{t-1}]\right\} &
			\mathbb{E}[Z_{t-1}'] \\
			\left\{-\frac{\beta_L}{\beta_0^2}(1-\rho_1)\mathbb{E}[\breve{m}_{t-1}]+\frac{\delta_{LM}}{\beta_0^2}\left[\mathbb{E}[\breve{m}_{t-1}S^L_{t}]-\rho_1\mathbb{E}[\breve{m}_{t-1}S^L_{t-1}]\right]\right\} &
			\frac{1-\rho_1}{\beta_0}\mathbb{E}[\breve{m}_{t-1}] &
			\left\{\mathbb{E}[\breve{m}_{t-1}^2] +\frac{\beta_L}{\beta_0}\mathbb{E}[\breve{m}_{t-1}]-\frac{\delta_{LM}}{\beta_0} \mathbb{E}[\breve{m}_{t-1}S^L_{t-1}]\right\} &
			\mathbb{E}[\breve{m}_{t-1}Z_{t-1}'] \\
			\left\{-\frac{\beta_L}{\beta_0^2}(1-\rho_1)\mathbb{E}[S^L_{t-1}]+\frac{\delta_{LM}}{\beta_0^2}\left[\mathbb{E}[S^L_{t-1}S^L_{t}]-\rho_1\mathbb{E}[(S^L_{t-1})^2]\right]\right\} &
			\frac{1-\rho_1}{\beta_0}\mathbb{E}[S^L_{t-1}] &
			\left\{\mathbb{E}[\breve{m}_{t-1}S^L_{t-1}] +\frac{\beta_L}{\beta_0}\mathbb{E}[S^L_{t-1}]-\frac{\delta_{LM}}{\beta_0} \mathbb{E}[(S^L_{t-1})^2]\right\} &
			\mathbb{E}[S^L_{t-1}Z_{t-1}'] \\
			\left\{-\frac{\beta_L}{\beta_0^2}(1-\rho_1)\mathbb{E}[Z_{t-1}]+\frac{\delta_{LM}}{\beta_0^2}\left[\mathbb{E}[S^L_{t}Z_{t-1}]-\rho_1\mathbb{E}[S^L_{t-1}Z_{t-1}]\right]\right\} &
			\frac{1-\rho_1}{\beta_0}\mathbb{E}[Z_{t-1}] &
			\left\{\mathbb{E}[\breve{m}_{t-1}Z_{t-1}] +\frac{\beta_L}{\beta_0}\mathbb{E}[Z_{t-1}]-\frac{\delta_{LM}}{\beta_0} \mathbb{E}[S^L_{t-1}Z_{t-1}]\right\} &
			\mathbb{E}[Z_{t-1}Z_{t-1}']
		\end{pmatrix}
	\end{longeq}}
where $\breve{m}_{t-1}=m_{t-1}-l_{t-1}$ is the logged material-to-labor ratio.


\section{Semiparametric Sieve Estimation}
\label{sec:appx_semi}

In what follows, we describe how to empirically implement our methodology semiparametrically, with the unknown functions $r_{\varphi}(\cdot)$ and $r_{\omega}(\cdot)$ approximated using linear sieves. Sieves globally approximate unknown nonparametric (i.e., infinite-parameter) functions using a sequence of less complex parameter spaces that are characterized by a finite number of ``parameters'' which effectively reduces the estimation problem to a parametric estimation when implemented in practice, with the caveat being that the complexity of such an approximation increases with the sample size to ensure consistency. For more on sieves, see an excellent review by \citet{chen2007}. 

The first-step estimation remains the same as described in Section \ref{sec:estimation}. In the second step, however, we now approximate the unknown function $r_{\varphi}(\varphi_{it-1},Z_{it-1})$ using a linear series of $[\dim(Z)+1]$-variate polynomial basis functions $\{\mathscr{P}_{r,n}(\varphi_{it-1},Z_{it-1})\}_{r=1}^{R_n}$ without the intercept term, with the degree of approximation complexity $R_n \to\infty$ slowly with $n$. As before, the unobservable $\varphi_{it-1}$ is replaced with a proxy function from \eqref{eq:lmratio_varphi}, and $\delta_{LM}=\beta_L+\beta_M$ is replaced with its estimate from the first step. Modifying the GMM problem in \eqref{eq:sst_ident_gmm_est} to accommodate a series approximation of $r_{\varphi}(\cdot)$, the second-step equation is now estimated for a given $R_n$ via semiparametric nonlinear sieve GMM. Thus, letting $\alpha=(\beta_0,\beta_L,\rho_{1},\dots,\rho_{R_n})'$, we have that
\begin{align}\label{eq:sst_ident_gmm_est_sieve}
\widehat{\alpha} = \arg\min_{\alpha} \Big[\tfrac{1}{nT}\sum_{it}\mathbb{Q}_{it-1}\epsilon_{it}(D_{it};\alpha)\Big]'W\Big[\tfrac{1}{nT}\sum_{it}\mathbb{Q}_{it-1}\epsilon_{it}(D_{it};\alpha)\Big],
\end{align}
where the approximated residual function is 
\begin{align}
\epsilon_{it}(D_{it};\alpha)\approx m_{it}-l_{it} +\frac{\beta_L}{\beta_0}-\frac{\widehat{\delta}_{LM}}{\beta_0} S^L_{it} - \sum_{r= 1}^{R_n}\rho_{r}\mathscr{P}_{r,n}\left(\left[m_{it-1}-l_{it-1} +\frac{\beta_L}{\beta_0}-\frac{\widehat{\delta}_{LM}}{\beta_0} S^L_{it-1}\right],Z_{it-1}\right),
\end{align}
and the instrument vector $\mathbb{Q}_{it-1}$ now needs to include not only the linear terms $(m_{it}-l_{it},Z_{it-1},S_{it-1}^L)'$ but also the additional higher-order terms from a polynomial expansion of $(m_{it}-l_{it},Z_{it-1},S_{it-1}^L)'$.

Then, just like in a fully parametric case, with the $\big(\widehat{\beta}_0,\widehat{\beta}_L\big)'$ estimates in hand, we can recover $\widehat{\beta}_M=\widehat{\delta}_{LM}-\widehat{\beta}_L$ as well as estimate Harrod-neutral productivity as $\widehat{\varphi}_{it}=m_{it}-l_{it} +\widehat{\beta}_L/\widehat{\beta}_0-\widehat{\delta}_{LM}/\widehat{\beta}_0\times S^L_{it}$.

The smoothing parameter $R_n$ can be controlled indirectly by selecting the optimal degree of polynomial expansion $d_n$ via generalized cross-validation of \citet{cravenwahba1979}: \begin{equation}\label{eq:gcv}
d_n^* = \arg\min_{d_n} \frac{\tfrac{1}{nT}||(\text{I}_{nT}-{\Pi}_{R_n})\widehat{\boldsymbol{\varphi}}||^2}{\left[1-\tfrac{1}{nT}\text{tr}\{{\Pi}_{R_n}\}\right]^2}\quad\text{with}\ R_n=R_n(d_n),
\end{equation}
where, for a given $(\beta_0,\beta_L)'$, ${\Pi}_{R_n}= \mathbb{P}_{R_n}(\mathbb{P}_{R_n}'\mathbb{P}_{R_n})^{-1}\mathbb{P}_{R_n}'$ is a projection matrix defined using the matrix of basis functions $\mathbb{P}_{R_n}$ constructed by stacking up $P_{R_n}(\widehat{\varphi}_{it-1},Z_{it-1})= [\mathscr{P}_{1,n}(\widehat{\varphi}_{it-1},Z_{it-1}),\dots,$ $\mathscr{P}_{R_n,n}(\widehat{\varphi}_{it-1},Z_{it-1})]'$ in the ascending order of index $i$ first then index $t$. The column vector $\widehat{\boldsymbol{\varphi}}$ is stacked up similarly but using $\{\widehat{\varphi}_{it}\}$.

To estimate the third-step equation in \eqref{eq:tst}, we construct estimators of $y_{it}^*$ and $m_{it}^*$ using the results from steps one and two: $\widehat{y}_{it}^*= y_{it} - \widehat{\beta}_Mm_{it}-\widehat{\beta}_L[\widehat{\varphi}_{it}+l_{it}] 
+\tfrac{1}{2}\widehat{\beta}_0[m_{it}-\widehat{\varphi}_{it}-l_{it}]^2$ and $\widehat{m}^*_{it}=\ln \left[ P_{t}^M/P_{t}^Y\right]-\ln\widehat{\theta}- \ln\big(\widehat{\beta}_M-\widehat{\beta}_0[m_{it}-\widehat{\varphi}_{it}-l_{it}] \big) +(1-  \widehat{\beta}_M)m_{it}-\widehat{\beta}_L[\widehat{\varphi}_{it}+l_{it}] +\tfrac{1}{2}\widehat{\beta}_0[m_{it}-\widehat{\varphi}_{it}-l_{it}]^2$.
Then, approximating unknown $r_{\omega}(\cdot)$ using $[\dim(X)+1]$-variate polynomial sieves of degree $d'_{n}$, i.e.,
\begin{align}
r_{\omega}(\omega_{it-1},X_{it-1}) &\approx \sum_{r'= 1}^{R'_n}{\pi}_{r'}\mathscr{P}_{r'}\left(\big[\widehat{m}^*_{it-1}-\beta_Kk_{it-1}-\tfrac{1}{2}\beta_{KK}k_{it-1}^2\big],X_{it-1}\right), 
\end{align}
where $\omega_{it-1}$ is replaced with its proxy and $R'_n(d'_n)$ increases with the sample size, we estimate $\gamma=(\beta_K,\beta_{KK},$ $\pi_{1},\dots,\pi_{R'_n})'$ via semiparametric nonlinear sieve least squares in line with \eqref{eq:tst_nls_est}:
\begin{align}\label{eq:tst_nls_est_semi}
\widehat{\gamma} = \arg\min_{\gamma} \sum_{it}\Bigg[\widehat{y}_{it}^* - \beta_Kk_{it}-\tfrac{1}{2}\beta_{KK}k_{it}^2- \sum_{r'= 1}^{R'_n}{\pi}_{r'}\mathscr{P}_{r'}\left(\big[\widehat{m}^*_{it-1}-\beta_Kk_{it-1}-\tfrac{1}{2}\beta_{KK}k_{it-1}^2\big],X_{it-1}\right)\Bigg]^2.
\end{align}
Analogous to the second step above, $R'_n$ can be cross-validated indirectly by selecting the optimal degree of polynomial expansion $d'_n$ via generalized cross-validation.

Using the obtained $(\widehat{\beta}_K,\widehat{\beta}_{KK})'$ estimates, we then can construct the estimates of Hicks-neutral productivity as $\widehat{\omega}_{it} = y_{it} - \widehat{\beta}_Kk_{it}-\tfrac{1}{2}\widehat{\beta}_{KK}k_{it}^2-  \widehat{\beta}_Mm_{it}-\widehat{\beta}_L[\widehat{\varphi}_{it}+l_{it}] 
+\tfrac{1}{2}\widehat{\beta}_0[m_{it}-\widehat{\varphi}_{it}-l_{it}]^2-\widehat{\eta}_{it} $ using $\widehat{\eta}_{it}$ from step one.

\textsl{Inference.}\textemdash Due to a multi-step nature of our methodology and the presence of nonparametric components, computation of the asymptotic variance of the semiparametric estimator above is not that simple. For statistical inference in this case, we therefore use bootstrap; the algorithm is described in Appendix \ref{sec:appx_boot}. 


\section{Bootstrap Inference}
\label{sec:appx_boot}

We approximate the sampling distribution of our estimator via wild residual block bootstrap that takes into account a panel structure of the data as well as a sequential nature of our multi-step estimation procedure. Concretely, the bootstrap algorithm is as follows.
\begin{enumerate}\itemsep 0pt
    \item Compute the three steps of our estimation procedure using the original data. Denote the obtained point estimates of all the parameters and functions using a ``hat.'' Correspondingly, let the (negative of) first-step residuals be $\{\widehat{\eta}_{it}\}$, the second-step residuals be $\{\widehat{ \zeta}_{\varphi,it}\}$,  and the third-step residuals be $\{\widehat{ \zeta_{\omega,it}+ \eta_{it}}\} $. Recenter these.

	\item Generate bootstrap weights $\xi_{i}^b$ for all cross-sectional units $i=1,\dots,n$ from the \citet{mammen1993} two-point mass distribution:
	\begin{eqnarray}\label{eq:two_point}
		\xi_{i}^b=%
		\begin{cases}
			\frac{1+\sqrt{5}}{2} & \text{with prob.}\quad \frac{\sqrt{5}-1}{2\sqrt{5}} \\
			\frac{1-\sqrt{5}}{2} & \text{with prob.}\quad \frac{\sqrt{5}+1}{2\sqrt{5}}.
		\end{cases}
	\end{eqnarray}
	Next, for each observation $(i,t)$ with $i=1,\dots,n$ and $t=1,\dots,T$, jointly generate a new bootstrap first-step disturbance $\eta_{it}^b = \xi_{i}^b\times\widehat{\eta}_{it}$, a new bootstrap second-step disturbance $\zeta_{\varphi,it}^b=\xi_{i}^b\times\widehat{ \zeta}_{\varphi,it}$, and a new bootstrap third-step disturbance  $(\zeta_{\omega,it}+ \eta_{it})^b=\xi_{i}^b\times(\widehat{ \zeta_{\omega,it}+ \eta_{it}})$.
	
	\item Generate a new bootstrap first-step outcome variable via $\ln R_{it} ^b=\ln \left[\widehat{\theta}(\widehat{\beta}_M+ \widehat{\beta}_L)\right]-\eta_{it}^b$ for all $i=1,\dots,n$ and $t=1,\dots,T$.  
	
	\item Generate a new bootstrap second-step outcome variable recursively as $(m_{it}- l_{it})^{b}=-\frac{\widehat{\beta}_L}{\widehat{\beta}_0}+\left(\frac{\widehat{\beta}_L+\widehat{\beta}_M}{\widehat{\beta}_0}\right) S^L_{it}+\widehat{r}_{\varphi,1}  \left[(m_{it-1}- l_{it-1})^{b} +\frac{\widehat{\beta_L}}{\widehat{\beta_0}}-\left(\frac{\widehat{\beta}_L+\widehat{\beta}_M}{\beta_0}\right) S^L_{it-1}\right]+ \widehat{r}_{\varphi,2} Z_{it-1}+\zeta_{\varphi,it}^b $  for all $i=1,\dots,n$ and $t=1,\dots,T$. To initialize at $t=0$, we set $(m_{i0}- l_{i0})^{b} = (m_{i0}- l_{i0})$.

    \item Generate a new bootstrap third-step outcome variable via
    $y_{it}^{*b}=\widehat{\beta}_K k_{it}+  \frac{1}{2}\widehat{\beta}_{KK} k^2_{it} + \widehat{r}_{\omega,0}+ \widehat{r}_{\omega,1} \Big[\widehat{m}^{*}_{it-1}-\widehat{\beta}_K k_{it}-  \frac{1}{2}\widehat{\beta}_{KK} k^2_{it} \Big]+\widehat{r}_{\omega,2}X_{it-1}+ (\zeta_{\omega,it}+ \eta_{it})^b$ for all $i=1,\dots,n$ and $t=1,\dots,T$. 
    
    \item Recompute the first step using $\{\ln R_{it} ^b\}$ in place of $\{\ln R_{it} \}$. Denote the obtained parameter estimates as $\big((\widehat{\beta}_M+  \widehat{\beta}_L)^b,\widehat{\theta}^b\big)'$.  
	
	\item Recompute the second step using $\{(m_{it}- l_{it})^{b}\}$ in place of $\{(m_{it}- l_{it})\}$ and using  $\{(m_{it-1}- l_{it-1})^{b}\}$ in place of $\{(m_{it-1}- l_{it-1})\}$. Denote the obtained parameter and function estimates as $\big(\widehat{\beta}_M^b,  \widehat{\beta}_L^b,\widehat{\beta}^b_0,\widehat{r}_{\varphi}(\cdot)\big)'$. Also obtain $ \widehat{\varphi}^b_{it}= m_{it}- l_{it}+\frac{\widehat{\beta}^b_L}{\widehat{\beta}^b_0}-\left(\frac{\widehat{\beta}^b_L+\widehat{\beta}^b_M}{\widehat{\beta}^b_0}\right) S^L_{it}$.
	
	\item Recompute the third step using $y_{it}^{*b}$ in place of $y_{it}^{*}$.  When re-estimating the equation, also use	$m_{it}^{*b}$  in place of 	$m_{it}^{*}$,  where 	$m_{it}^{*b}=\ln \left[ \frac{P_{t}^M}{P_{t}^Y}\right]-\ln \theta^b- \ln\big(\widehat{\beta}^b_M-\widehat{\beta}^b_0[m_{it}-\widehat{\varphi}^b_{it}-l_{it}] \big) +(1-  \widehat{\beta}^b_M)m_{it}-\widehat{\beta}_L[\widehat{\varphi}^b_{it}+l_{it}] +\tfrac{1}{2}\widehat{\beta}^b_0[m_{it}-\widehat{\varphi}^b_{it}-l_{it}]^2 $ for all $i=1,\dots,n$ and $t=1,\dots,T$. Denote the obtained parameter and function estimates as  $\big(\widehat{\beta}_K^b,\widehat{\beta}_{KK}^b,\widehat{r}^b_{\omega}(\cdot))'$.

	\item Repeat steps 2 through 8 of the algorithm $B$ times.
\end{enumerate} 

Let the estimand of interest be denoted by $\mathcal{E}$, e.g., the firm $i$'s capital elasticity $\epsilon_{K,it}\equiv\beta_K+\beta_{KK}k_{it}$ at time $t$. To perform hypothesis testing, we can use the empirical distribution of $\{\widehat{\mathcal{E}}^1,\dots,\widehat{\mathcal{E}}^B\}$ to obtain a bootstrap estimator of $\mathbb{V}\big[\widehat{\mathcal{E}}\big]$ following the 
standard methods.

\section{Data Summary}
\label{sec:appx_data}

\setcounter{figure}{0} \renewcommand\thetable{\thesection.\arabic{table}}
\setcounter{table}{0} \renewcommand\thefigure{\thesection.\arabic{figure}}

\begin{table}[h]
    \caption{Data Summary Statistics}
    \label{tab:data summary}
  \centering \small
  \makebox[\linewidth]{
  \begin{tabular}{lrrrrrrrrr}
\toprule[1pt]
                         & \multicolumn{4}{c}{$Z = 0$ \ (Obs \#: 27,094)}          & \multicolumn{1}{l}{} & \multicolumn{4}{c}{$Z > 0$ \ (Obs \#: 4,193)} \\ 
                         & Mean  & 1st Qu. & Median & 3rd Qu. &                      & Mean   & 1st Qu.  & Median & 3rd Qu. \\
                         \cline{2-5} \cline{7-10} 
Output ($Y$)               & 29,719 & 10,554   & 19,353  & 37,684   &                      & 38,611  & 14,483    & 28,312  & 52,405   \\
Capital ($K$)              & 3,863  & 926     & 2,215   & 4,877    &                      & 4,933   & 1,296     & 3,003   & 6,534    \\
Labor ($L$)                & 2,140  & 729     & 1,370   & 2,659    &                      & 3,069   & 1,199     & 2,181   & 4,018    \\
Materials ($M$)             & 20,448 & 7,159    & 13,251  & 25,941   &                      & 26,833  & 9,665     & 19,220  & 36,612   \\
Foreign equity share ($Z$) &       &         &        &         &                      & 0.47   & 0.21     & 0.41   & 0.72    \\ 
		\midrule
\multicolumn{10}{p{15.5cm}}{\footnotesize {\sc Notes:} The three inputs and the output are in thousands of real RMB. The foreign equity share is a unit-free proportion.} \\ \bottomrule[1pt]
\end{tabular}
  }
\end{table}


\section{Additional Empirical Results}
\label{sec:appx_emprobust}

\setcounter{figure}{0} \renewcommand\thetable{\thesection.\arabic{table}}
\setcounter{table}{0} \renewcommand\thefigure{\thesection.\arabic{figure}}

The two tables below report the estimates of the production function and productivity parameters obtained when using the inverted conditional labor demand in \eqref{eq:laborproxy} [Table \ref{tab:prod fn parameter_lstar}] and the average of the inverted conditional material and labor demands in \eqref{eq:mdemand_omega} and \eqref{eq:laborproxy} [Table \ref{tab:prod fn parameter_avemlstar}] to proxy for latent $\omega_{it}$. Since this proxy is used in the third step of our estimator, only the estimates of capital elasticity and the Hicks-neutral productivity process are affected; the remaining parameters are the same as those reported in Table \ref{tab:prod fn parameter}.

\medskip

\begin{table}[h]
	\centering \small
	\caption{Estimates of the Production Function and Productivity Parameters using the Labor Proxy}
	\label{tab:prod fn parameter_lstar}
	\begin{tabular}{ccccc}
		\toprule[1pt]
		\multicolumn{5}{c}{\bf Panel A: Elasticities and Returns to Scale}                                                                     \\
		& Mean     & 1st Qu.  & Median                               & 3rd Qu.  \\ \cline{2-5} 
		\multicolumn{1}{l}{Capital elasticity}  & 0.0436   & 0.0283   & 0.0443                               & 0.0587   \\
		& (0.0053) & (0.0067) & (0.0053)                             & (0.0080) \\
		\multicolumn{1}{l}{Labor elasticity}    & 0.0950   & 0.0458   & 0.0864                               & 0.1287   \\
		& (0.0003) & (0.0001) & (0.0002)                             & (0.0004) \\
		\multicolumn{1}{l}{Material elasticity} & 0.7377   & 0.7041   & 0.7463                               & 0.7869   \\
		& (0.0020) & (0.0019) & (0.0021)                             & (0.0022) \\
		\multicolumn{1}{l}{RTS}                 & 0.8763   & 0.8610   & 0.8771                               & 0.8915   \\
		& (0.0057) & (0.0071) & (0.0058)                             & (0.0083) \\[4pt]
		\midrule
		\multicolumn{5}{c}{\bf Panel B: Productivity Parameters}  \\ 
		\multicolumn{2}{c}{\sl Labor-Augmenting} & & \multicolumn{2}{c}{\sl Factor-Neutral} \\ \cline{1-2} \cline{4-5}
		\multicolumn{1}{l}{$\rho_{\varphi,0}$}& 0.0000 &&\multicolumn{1}{l}{$\rho_{\omega,0}$} & 0.8082   \\
		&   ---    &          &                                      & (0.0928) \\
		\multicolumn{1}{l}{$\rho_{\varphi,1}$}   & 0.5723   &          & \multicolumn{1}{l}{$\rho_{\omega,1}$} & 0.6840   \\
		& (0.0244) &          &                                      & (0.0247)  \\
		\multicolumn{1}{l}{$\rho_{\varphi,2}$}   & 0.0726   &          & \multicolumn{1}{l}{$\rho_{\omega,2}$} & 0.0045    \\
		& (0.0183) &          &                                      & (0.0084)  \\[4pt] 
		\midrule
		\multicolumn{5}{p{9cm}}{\footnotesize {\sc Notes:} The productivity processes are parameterized as follows:  $\varphi_{it}=\rho_{\varphi,1}\varphi_{it-1}+\rho_{\varphi,2}Z_{it-1}+\zeta_{\varphi,it}$ with $\rho_{\varphi,0}$ normalized to 0, and $\omega_{it}=\rho_{\omega,0}+\rho_{\omega,1}\omega_{it-1}+\rho_{\omega,2}Z_{it-1}+\zeta_{\omega,it}$. Bootstrap standard errors are in parentheses.} \\
		\bottomrule[1pt]
	\end{tabular}
\end{table}%

\clearpage
\begin{table}[h]
	\centering \small
	\caption{Estimates of the Production Function and Productivity Parameters using the Average of Material and Labor Proxies}
	\label{tab:prod fn parameter_avemlstar}
	\begin{tabular}{ccccc}
		\toprule[1pt]
		\multicolumn{5}{c}{\bf Panel A: Elasticities and Returns to Scale}                                                                     \\
		& Mean     & 1st Qu.  & Median                               & 3rd Qu.  \\ \cline{2-5} 
		\multicolumn{1}{l}{Capital elasticity}  & 0.0392   & 0.0243   & 0.0399                               & 0.0539   \\
		& (0.0065) & (0.0077) & (0.0066)                             & (0.0089) \\
		\multicolumn{1}{l}{Labor elasticity}    & 0.0950   & 0.0458   & 0.0864                               & 0.1287   \\
		& (0.0003) & (0.0001) & (0.0002)                             & (0.0004) \\
		\multicolumn{1}{l}{Material elasticity} & 0.7377   & 0.7041   & 0.7463                               & 0.7869   \\
		& (0.0020) & (0.0019) & (0.0021)                             & (0.0022) \\
		\multicolumn{1}{l}{RTS}                 & 0.8719   & 0.8570   & 0.8726                               & 0.8866   \\
		& (0.0068) & (0.0080) & (0.0069)                             & (0.0091) \\[4pt]
		\midrule
		\multicolumn{5}{c}{\bf Panel B: Productivity Parameters}  \\ 
		\multicolumn{2}{c}{\sl Labor-Augmenting} & & \multicolumn{2}{c}{\sl Factor-Neutral} \\ \cline{1-2} \cline{4-5}
		\multicolumn{1}{l}{$\rho_{\varphi,0}$}& 0.0000 &&\multicolumn{1}{l}{$\rho_{\omega,0}$} & 0.7115   \\
		&   ---    &          &                                      & (0.0975) \\
		\multicolumn{1}{l}{$\rho_{\varphi,1}$}   & 0.5723   &          & \multicolumn{1}{l}{$\rho_{\omega,1}$} & 0.7277   \\
		& (0.0244) &          &                                      & (0.0279)  \\
		\multicolumn{1}{l}{$\rho_{\varphi,2}$}   & 0.0726   &          & \multicolumn{1}{l}{$\rho_{\omega,2}$} & 0.0033    \\
		& (0.0183) &          &                                      & (0.0083)  \\[4pt] 
		\midrule
		\multicolumn{5}{p{9cm}}{\footnotesize {\sc Notes:} The productivity processes are parameterized as follows:  $\varphi_{it}=\rho_{\varphi,1}\varphi_{it-1}+\rho_{\varphi,2}Z_{it-1}+\zeta_{\varphi,it}$ with $\rho_{\varphi,0}$ normalized to 0, and $\omega_{it}=\rho_{\omega,0}+\rho_{\omega,1}\omega_{it-1}+\rho_{\omega,2}Z_{it-1}+\zeta_{\omega,it}$. Bootstrap standard errors are in parentheses.} \\
		\bottomrule[1pt]
	\end{tabular}
\end{table}%


\section{Extension to Imperfect Competition}
\label{sec:appx_ext}

\setcounter{figure}{0} \renewcommand\thetable{\thesection.\arabic{table}}
\setcounter{table}{0} \renewcommand\thefigure{\thesection.\arabic{figure}}

Our methodology  and its underlying identification scheme in Sections \ref{sec:model}--\ref{sec:identification} are developed under the assumption that firms operate in a perfectly competitive output market. In this appendix, we discuss how to relax this assumption and allow for monopolistic power in the output market in our setup.

Let the output price be no longer $P^Y_{it}\ne P^Y_t\  \forall\ i$. To allow firms to have some market power, we assume they produce differentiated products and operate in a monopolistically competitive market. Let each firm face a downward-sloping (residual) inverse demand function of the following generic form:\footnote{More generally, the residual demand that a firm faces can also depend on its rivals' prices. While we assume this away, one may be able to account for such substitution effects by replacing rivals' prices with an aggregate price index or dummies, although it may substantially increase the number of parameters to be estimated.} $P_{it}^Y=D(Y_{it}^e,U_{it})$, where $Y_{it}^e=Y_{it}\exp\{-\eta_{it}\}$ is the expected, or ``planned,'' output quantity net of an unanticipated \textit{ex-post} productivity shock $\eta_{it}$, and $U_{it}$ is a vector of demand shifters known to the firm at time $t$, i.e., $U_{it}\in\Xi_{it}$.\footnote{We can also allow for random price/demand shocks by, say, augmenting the demand equation with a log-additive unanticipated \textit{i.i.d.}~shock akin to $\eta_{it}$: $P_{it}^Y=D(Y_{it}^e,U_{it})\exp\{\varepsilon_{it}\}$ with $\mathbb{E}[\varepsilon_{it}|\Xi_{it}]=\mathbb{E}[\varepsilon_{it}]=0$. This would only result in an additional multiplicative constant entering the firm's static first-order conditions \textit{in expectation} [see eqs.~\eqref{eq:foc_l_moncomp}--\eqref{eq:foc_m_moncomp}] thereby having no material impact on the analysis.} But given our discussion concerning the oft-problematic use of firm-level variation in input prices for identification in Section \ref{sec:unidentification_standard}, we retain the assumption of competitive homogeneous factor prices.

In what follows, we fist show \textit{point} non-identification of the production function in \eqref{eq:prodfn_tl} when firms exhibit market power even if one observes exogenous demand shifters in the data. We then discuss how, with some additional but quite reasonable assumptions about unobservables, one can still \textit{set} identify the production function, and we derive this result without requiring that the information on demand shifters be available: we only  rely on the same set of observables that are used in our main methodology for the case of perfectly competitive markets from Section \ref{sec:identification}.


\subsection*{Point Non-Identification of the Production Function}

For a risk-neutral firm with market power, the firm's static optimality conditions are now given by
\begin{align}
	\left(1+\frac{1}{\delta(P^Y_{it},U_{it})}\right)P_{it}^Y \frac{\exp\{\overline{y}_{it}\}}{L_{it}}\big(\beta_L+\beta_0[m_{it}-\varphi_{it}-l_{it}] \big)\exp\{\omega_{it}\}\theta &=P_{t}^L, \label{eq:foc_l_moncomp} \\
	\left(1+\frac{1}{\delta(P^Y_{it},U_{it})}\right)P_{it}^Y \frac{\exp\{\overline{y}_{it}\}}{M_{it}}\big(\beta_M-\beta_0[m_{it}-\varphi_{it}-l_{it}] \big)\exp\{\omega_{it}\}\theta &=P_{t}^M, \label{eq:foc_m_moncomp}
\end{align}
where $\delta(P^Y_{it},U_{it})<-1$ is the price elasticity of demand. The ratio of these optimality conditions however remains unchanged [see eq.~\eqref{eq:lmratio}] implying that the proxy for Harrod-neutral productivity $\varphi_{it}$ is also unchanged [see eq.~\eqref{eq:lmratio_varphi}]. But the new proxy for Hicks-neutral productivity does need to explicitly account for the firm's market power (here we continue to use the inverted material demand):
\begin{align}\label{eq:mdemand_omega_moncomp}
	\omega_{it} =&\ \overbrace{\ln \left[ \frac{P_{t}^M}{P_{it}^Y}\right]-\ln\theta- \ln\big(\beta_M-\beta_0[m_{it}-\varphi_{it}-l_{it}] \big) +(1-  \beta_M)m_{it}-\beta_L[\varphi_{it}+l_{it}] 
		+\tfrac{1}{2}\beta_0[m_{it}-\varphi_{it}-l_{it}]^2}^{m^*_{it}}
	\notag \\
	&\  -\beta_Kk_{it}-\tfrac{1}{2}\beta_{KK}k_{it}^2 + \ln \mu(P^Y_{it},U_{it}),
\end{align}
where $\mu(P^Y_{it},U_{it})=\left( 1+\frac{1}{\delta(P^Y_{it},U_{it})} \right)^{-1}$ is a markup (a price-to-marginal-cost ratio).

Now, consider the first step of our methodology. Because the new first-order conditions in \eqref{eq:foc_l_moncomp}--\eqref{eq:foc_m_moncomp} contain the demand elasticity, the variable input share equations will also have to account for markups:
\begin{align}
	\ln V_{it}^L &= \ln \left( \theta\beta_0\left[\frac{\beta_L}{\beta_0}+ m_{it}-\varphi_{it}-l_{it}\right] \right) - \ln \mu(P^Y_{it},U_{it})  - \eta_{it}, \label{eq:share_l_moncomp} \\
	\ln V_{it}^M &= \ln \left( \theta\beta_0\left[\frac{\beta_M}{\beta_0}- m_{it}+\varphi_{it}+l_{it}\right] \right) -\ln \mu(P^Y_{it},U_{it}) - \eta_{it}. \label{eq:share_m_moncomp}
\end{align}

Controlling for $\varphi_{it}$ using the (unchanged) material-to-labor ratio proxy function in \eqref{eq:lmratio_varphi}, we thus obtain the following variable-input-cost-to-revenue equation for a monopolistically competitive firm:
\begin{align}\label{eq:fst_moncomp}
	\ln R_{it} &= \ln \big( \theta\left[\beta_L+\beta_M\right] \big) - \ln \mu(P^Y_{it},U_{it}) - \eta_{it},
\end{align}
in which all regressors are weakly exogenous with respect to an \textit{ex-post} shock $\eta_{it}$.

The two components $\ln \big( \theta\left[\beta_L+\beta_M\right] \big)$ and $\ln \mu(P^Y_{it},U_{it})$ in \eqref{eq:fst_moncomp} are however not \textit{separably} identified. The unknown function $\mu(\cdot)$ can be identified up to a scale only. With the lack of point identification of the firm's markup $\mu(P^Y_{it},U_{it})$, the production-function parameters $(\beta_L,\beta_M)'$ cannot be point identified either.\footnote{$\theta$ is point identified because the random shocks $\{\eta_{it}\}$ are point identified via $\eta_{it}= \ln \big( \frac{\theta\left[\beta_L+\beta_M\right]}{\mu(P^Y_{it},U_{it})} \big)-\ln R_{it} $, where $\ln \big( \frac{\theta\left[\beta_L+\beta_M\right]}{\mu(P^Y_{it},U_{it})} \big)=\mathbb{E}[\ln R_{it}|P^Y_{it},U_{it}]$ is some unknown conditional mean function of $\ln R_{it}$ estimatable via least squares.} To see this more clearly, consider the special case of isoelastic demand whereby $\mu(P^Y_{it},U_{it})=\mu \ \forall \ i,t$. In this case, we can only point identify $(\beta_L+\beta_M)/\mu$ from \eqref{eq:fst_moncomp} and, consequently, we cannot point identify $(\beta_0,\beta_L)'$ from \eqref{eq:sst} either, with the similar implications for $(\beta_K,\beta_{KK})'$ in the third step. Thus, all production-function parameters can only be point identified up to a scale of the firm's markup $\mu$. More generally, the point non-identification of production function when firms exercise monopolistic power (although when productivity is uni-dimensional) is also discussed in \citet{flynnetal2019}.

Interestingly, despite the lack of point identification of both the production-function parameters and the markup, the variable-input-cost-to-revenue equation under imperfect competition in \eqref{eq:fst_moncomp} may nonetheless provide some useful information about \textit{markups} if $P^Y_{it}$ and $U_{it}$ are observed in the data. Namely, oftentimes markups \textit{per se} are of little policy relevance and, instead, economists focus on their relation with some other correlates or their distribution across firms. For instance, one might be interested in testing if exporters enjoy a greater price-setting power that do wholly domestically oriented firms \citep[e.g.,][]{,deloeckerwarzynski2012,deloeckeretal2016}. Alternatively, we may be interested in temporal dynamics of markups \citep[e.g.,][]{deloeckeretal2020,deloeckereeckhour2020}, be it on average or in terms of their cross-firm dispersion. Such analyses of markups are customarily done by regressing the estimated (log) markups on the variables of interest such export status or time trend/dummies. We can still accomplish the latter using ``scaled'' markup estimates from \eqref{eq:fst_moncomp}. Namely, $\ln \big( \tfrac{\theta\left[\beta_L+\beta_M\right]}{\mu(P^Y_{it},U_{it})} \big)$ in \eqref{eq:fst_moncomp} is some unknown function of $(P^Y_{it},U_{it}')'$ easily estimable via nonparametric least squares by regressing $\ln R_{it}$ on $(P^Y_{it},U_{it}')'$ with intercept. Then, so long as the production function is correctly specified and thus $\beta_L+\beta_M$ is a constant, we can regress the recovered scaled log-markup $\ln \big( \tfrac{\mu(P^Y_{it},U_{it})}{\theta\left[\beta_L+\beta_M\right]} \big)=-\ln \big( \tfrac{\theta\left[\beta_L+\beta_M\right]}{\mu(P^Y_{it},U_{it})} \big)$ on whatever variables of interest, with the only (and unimportant) implication being a bias in the intercept. Analogously, we can study the dispersion in markups or changes therein over time by analyzing the ratios of $\tfrac{\mu(P^Y_{it},U_{it})}{\theta\left[\beta_L+\beta_M\right]}$ across firms or the shifts in their distributions over time.


\subsection*{Partial Identification of the Production Function}

Although the production function is not point-identified when firms have monopolistic power in the output markets, we can still achieve its partial identification. To set identify the production-function parameters $(\beta_K,\beta_{KK},\beta_M,\beta_L,\beta_0)'$, we build upon \citeauthor{demirer2019}'s (2019) framework which we modify to admit multi-dimensional firm productivity. 

Since most production datasets contain no information on exogenous firm demand shifters, here we can also relax the assumption that demand heterogeneity $U_{it}$ be observable. But with the introduction of new unobservables, we need to formalize their relation to other unobservables that are known to the firm such as productivity. Defining the vector of \textit{observable} (by an econometrician) state variables as $O_{it}=(K_{it},P_{t}^L,P_{t}^M,X_{it}',Z_{it}')'$, we augment our assumptions as follows.

\begin{assumption}[\normalfont Replaces Assumption \ref{assm:4}]\label{assm:5} {\normalfont (i)} Risk-neutral firms maximize the discounted stream of life-time profits in perfectly competitive factor markets with homogeneous prices. The output market is monopolistically competitive, and the firm's downward-sloping inverse (residual) demand function is given by $P_{it}^Y=D(Y_{it}^e,U_{it})$ and $U_{it}\in\Xi_{it}$ is a vector of unobservable (to an econometrician) demand shifters known to the firm when making time $t$ decisions. {\normalfont (ii)}Conditional on observable $O_{it-1}$, demand heterogeneity $U_{it}$ is jointly independent of firm productivity $\varphi_{it}$ and $\omega_{it}$.\footnote{Assumption \ref{assm:5}(ii) can be relaxed by making the independence be conditional on $\varphi_{it-1}$ and $\omega_{it-1}$ thereby, in effect, assuming that $U_{it}$ is jointly independent of the contemporaneous \textit{innovations} in firm productivities.}
\end{assumption}

In the above, $Y_{it}^e=Y_{it}\exp\{-\eta_{it}\}$ is the expected/planned output quantity as defined earlier. Since the presence of markups does not in any way affect the ratio of firm's static first-order conditions, thereby providing a deterministic proxy for labor-augmenting productivity given by \eqref{eq:lmratio_varphi}, we can utilize it to substitute latent $\varphi_{it}$ out of the production function  \eqref{eq:prodfn_tl} to arrive at
\begin{align}\label{eq:prodfn_tl_novarphi}
	y_{it} &= \beta_Kk_{it}+\tfrac{1}{2}\beta_{KK}k_{it}^2+  (\beta_L+\beta_M)m_{it}+ \tfrac{1}{2\beta_0}\left[\beta_L^2-(\beta_L+\beta_M)^2\big(S_{it}^L\big)^2\right]+
	\omega_{it}+\eta_{it} \notag \\ 
	&= \underbrace{\tfrac{\beta_L^2}{2\beta_0} +\beta_Kk_{it}+\tfrac{1}{2}\beta_{KK}k_{it}^2+  (\beta_L+\beta_M)m_{it}- \tfrac{(\beta_L+\beta_M)^2}{2\beta_0}\big(S_{it}^L\big)^2}_{\overline{y}_{it}(v_{it};\beta)} +\ 
	\omega_{it}+\eta_{it},
\end{align}
which now contains only \textit{additive} unobservables and where $v_{it}=(k_{it},m_{it},l_{it},P_{t}^M,P_t^L)'$ and $\beta=(\beta_K,\beta_{KK},$ $\beta_L,\beta_M,\beta_0)'$.\footnote{Recall that $S^L_{it}$ is a deterministic function of $P_{t}^M,P_t^L,\exp\{m_{it}\}$ and $\exp\{l_{it}\}$.} In what follows, we seek to set-identify $\beta$.

We first establish the relationship between flexible inputs and firm productivity, which facilitates the proxy variable approach to tackling unobservability of the latter. Consistent with our primary methodology, we use materials to control for productivity.

\setcounter{theorem}{0}
\begin{proposition}\label{prop:1}
	Under Assumptions \ref{assm:1}--\ref{assm:3} \& \ref{assm:5} and some additional regularization of the curvature of the production function and the firm's downward-sloping (residual) demand function, the firm's conditional demands for $M_{it}$ is weakly increasing in $\omega_{it}$ and $\varphi_{it}$, conditional on all other state variables entering the expected static profit maximization problem $(K_{it},P_{t}^L,P_{t}^M,U_{it}')'$.
\end{proposition}

The proposition signs partial derivatives of the material demand with respect to both components of firm productivity and is easy but tedious to show, which involves differentiation of first-order conditions in \eqref{eq:foc_l_moncomp}--\eqref{eq:foc_m_moncomp} with respect to productivities (see Appendix \ref{sec:appx_prop}). Intuitively, the firm (i) substitutes materials for labor conserving the latter \textit{ceteris paribus} as the labor input becomes more productive when $\varphi_{it}$ rises  and (ii) uses more materials (as well as labor) expanding the production when its overall total factor productivity $\omega_{it}$ improves increasing the marginal products of static inputs. For convenience, denote the material demand as $M_{it}=\mathcal{M}(K_{it},P_{t}^L,P_{t}^M,U_{it},\varphi_{it},\omega_{it})=\mathcal{M}(K_{it},U_{it},\varphi_{it},\omega_{it})$, where we suppress input prices because they provide little operationable information due to the lack of cross-firm variation under our assumptions.

To derive moment inequalities that partially identify the firm's production function, we need to tighten our assumptions by formalizing the relationship between the two components of firm productivity. As noted in Section \ref{sec:model}, Assumption \ref{assm:3}(i) places no restriction on the relation between $\omega_{it}$ and $\varphi_{it}$ but, to identify the production function when firms have market power, its now needs be regulated. We do so by letting the labor-augmenting productivity $\varphi_{it}$ be stochastically increasing in factor-neutral Hicksian productivity $\omega_{it}$, conditional on productivity ``controls.'' Thus, we extend our Assumption \ref{assm:3}(i) as follows.

\setcounter{theorem}{5}
\begin{assumption}\label{assm:6} Conditional on productivity-modifying controls $(X_{it-1}',Z_{it-1}')'$, the distribution of $\varphi_{it}$ is stochastically increasing in $\omega_{it}$.
\end{assumption}

More specifically, Assumption \ref{assm:6} means that $\mathcal{P}_{\varphi}(\varphi_{it}|\omega_{it}^H,X_{it-1},Z_{it-1})$ first-order stochastically dominates $\mathcal{P}_{\varphi}(\varphi_{it}|\omega_{it}^L,X_{it-1},Z_{it-1})$ iff $\omega_{it}^H\ge \omega_{it}^L$. In this, we intuitively assume that the firms that are more productive in general (in all factors) are likely to also be more productive in labor. Then, the set identification of the production-function parameters $\beta$ is obtained based on the following proposition, according to which the material proxy can be used to stochastically order the log-additive (and only) unobservable $\omega_{it}+\eta_{it}$ entering the production function in \eqref{eq:prodfn_tl_novarphi}.

\setcounter{theorem}{1}
\begin{proposition}\label{prop:2}
	For some cutoff value $\widetilde{\mathrm{m}}$ for the materials input, let $\mathbb{S}=\{(o,\widetilde{\mathrm{m}}): \Pr(m_{it}<\widetilde{\mathrm{m}}|O_{it-1}=o)\in(0,1)\}$ denote the common support. Under Assumptions \ref{assm:1}--\ref{assm:3} \& \ref{assm:5}--\ref{assm:6} and by Proposition \ref{prop:1}, for $(O_{it-1},\widetilde{\mathrm{m}})\in\mathbb{S}$ we have
	\begin{equation}\label{eq:prop}
		\mathbb{E}\left[y_{it}-\overline{y}(v_{it};\beta)|O_{it-1},m_{it}>\widetilde{\mathrm{m}}\right] \ge \mathbb{E}\left[y_{it}-\overline{y}_{it}(v_{it};\beta)|O_{it-1},m_{it}<\widetilde{\mathrm{m}}\right].
	\end{equation}
\end{proposition}

The proof is in Appendix \ref{sec:appx_prop}.  By this proposition, when comparing high-materials (those with $m_{it}>\widetilde{\mathrm{m}}$) and low-materials (those with $m_{it}<\widetilde{\mathrm{m}}$) firms, the firms that use more inputs are more Hicks-productive on average. In $y_{it}-\overline{y}(v_{it};\beta)=\omega_{it}+\eta_{it}$, it is evident that the focus here is on factor-neutral productivity without the need to characterize labor-augmenting productivity. This is possible owing to the availability of a deterministic proxy of the known functional form for $\varphi_{it}$ afforded to us by our parametric specification of the firm's separable production technology, which enable us to concentrate Harrod-neutral productivity from production model in \eqref{eq:prodfn_tl_novarphi}.

Conditional on $o$ and $\widetilde{\mathrm{m}}$, partially identified $\beta$ parameters are a nonlinear half-space, and the identified set $\mathcal{B}^*$ is the intersection of these half-planes:
\begin{equation}
	\mathcal{B}^* = \Big\{ \breve{\beta}\in\mathcal{B}: \cap_{(o,\widetilde{\mathrm{m}})\in\mathbb{S}} \mathbb{E}\left[y_{it}-\overline{y}(v_{it};\breve{\beta})|O_{it-1}=o,m_{it}>\widetilde{\mathrm{m}}\right] - \mathbb{E}\left[y_{it}-\overline{y}_{it}(v_{it};\breve{\beta})|O_{it-1}=o,m_{it}<\widetilde{\mathrm{m}}\right] \ge 0 \big\},
\end{equation}
which contains true $\beta\in\mathcal{B}^*$ and where $\mathcal{B}$ is a compact parameter space.

To operationalize this partial identification result, we can redefine the moment inequality in \eqref{eq:prop} using inverse propensity score weighting. Consider a binary variable $\mathbbm{1}\{m_{it}>\widetilde{\mathrm{m}}\}$ which delineates high- and low-materials firms for a given cutoff $\widetilde{\mathrm{m}}$ and which corresponds to the conditioning event of interest. Noting that $\Pr(m_{it}>\widetilde{\mathrm{m}}|O_{it-1})=\mathbb{E}[\mathbbm{1}\{m_{it}>\widetilde{\mathrm{m}}\}|O_{it-1}]$, we then have
\begin{align}
	\mathbb{E}\left[y_{it}-\overline{y}(v_{it};\beta)|O_{it-1},m_{it}>\widetilde{\mathrm{m}}\right]&=
	\mathbb{E}\left[\frac{\big(y_{it}-\overline{y}(v_{it};\beta)\big)\mathbbm{1}\{m_{it}>\widetilde{\mathrm{m}}\}}{\mathbb{E}[\mathbbm{1}\{m_{it}>\widetilde{\mathrm{m}}\}|O_{it-1}]}\Big|O_{it-1}\right] , \\ \mathbb{E}\left[y_{it}-\overline{y}_{it}(v_{it};\beta)|O_{it-1},m_{it}<\widetilde{\mathrm{m}}\right]&=
	\mathbb{E}\left[\frac{\big(y_{it}-\overline{y}(v_{it};\beta)\big)\big(1-\mathbbm{1}\{m_{it}>\widetilde{\mathrm{m}}\}\big)}{1-\mathbb{E}[\mathbbm{1}\{m_{it}>\widetilde{\mathrm{m}}\}|O_{it-1}]}\Big|O_{it-1}\right],
\end{align}
which we can further transform by integrating multi-dimensional $O_{it-1}$ out to arrive at the unconditional moment inequality:
\begin{equation}\label{eq:prop_uncond}
	\mathbb{E}\left[\frac{\big(y_{it}-\overline{y}(v_{it};\beta)\big)\mathbbm{1}\{m_{it}>\widetilde{\mathrm{m}}\}}{\mathbb{E}[\mathbbm{1}\{m_{it}>\widetilde{\mathrm{m}}\}|O_{it-1}]}\right] - 
	\mathbb{E}\left[\frac{\big(y_{it}-\overline{y}(v_{it};\beta)\big)\big(1-\mathbbm{1}\{m_{it}>\widetilde{\mathrm{m}}\}\big)}{1-\mathbb{E}[\mathbbm{1}\{m_{it}>\widetilde{\mathrm{m}}\}|O_{it-1}]}\right] \ge 0.
\end{equation}

Propensity scores $\Pr(m_{it}>\widetilde{\mathrm{m}}|O_{it-1})=\mathbb{E}[\mathbbm{1}\{m_{it}>\widetilde{\mathrm{m}}\}|O_{it-1}]$ can be estimated via one of many semi- or nonparametric estimators for binary outcomes. With this, a confidence set for true $\beta$ whose values are restricted by the moment inequality in \eqref{eq:prop_uncond} can be estimated by inverting a test corresponding to this moment condition. Essentially, one is to look for a set of $\beta$ values for which one fails to reject the null that the difference in means in \eqref{eq:prop_uncond} is positive. The literature on inference using moment inequalities (especially in industrial organization) is vast, and many moment inequality estimation and inference frameworks are readily available to be used here, e.g., see \citet{canayshaikh2017,molinari2020,klineetal2021,stoye2021} and many citations therein.


\section{Proofs}
\label{sec:appx_prop}

\subsection*{Proof of Proposition \ref{prop:1}}
We examine Proposition \ref{prop:1} under two scenarios: (i) a special case of the isoelastic demand function and (ii) a more general case when the demand elasticity is not constant. The idea of the proof for these two scenarios is the same, but the assumption of a constant elasticity of demand simplifies the mathematical derivation.

In what follows, the monotonicity of the firm's conditional material demand $\mathcal{M}(\cdot)$ with respect to the two components of firm productivity is derived under Assumptions \ref{assm:1}--\ref{assm:3} \& \ref{assm:5} \textit{and} assuming two additional regularity conditions on the curvature of the production function and the firm's downward-sloping (residual) demand function. Namely, we assume that (i) the cross-elasticities of variable inputs are non-negative, i.e., $\frac{\partial^2 \ln F(\cdot)}{\partial \ln L\partial \ln M} \ge 0$,\footnote{Intuitively, this is akin to a restriction that variable inputs be ``gross complements.''} and (ii) the price elasticity of markup is within the unit interval: $0 \leq \frac{\partial \ln \mu}{\partial \ln P^Y} \leq 1$.  

Let us first rewrite the firm's static optimality conditions as
\begin{align}\label{eq:foc_prop1}
    P^Y \frac{\partial Y^e}{\partial M} \theta \mu^{-1} = P^M  \quad \text{and} \quad   P^Y \frac{\partial Y^e}{\partial L} \theta \mu^{-1} = P^L,
\end{align}
where the individual and time subscripts are suppressed for the ease of notation. 

\textsl{Isoelastic demand.}\textemdash First, we show that the conditional material demand $\mathcal{M}(\cdot)$ is weakly increasing in $\omega$. Differentiating the first-order-condition equations in \eqref{eq:foc_prop1} with respect to $\omega$, we obtain
\begin{align*}
    &\underbrace{\left( \frac{\partial P^Y}{\partial Y^e} \left(\frac{\partial Y^e}{\partial M}\right)^2 + P^Y \frac{\partial^2 Y^e}{\partial M^2} \right)}_{a_1} \frac{\partial M}{\partial \omega} + \underbrace{\left( \frac{\partial P^Y}{\partial Y^e} \frac{\partial Y^e}{\partial M}\frac{\partial Y^e}{\partial L} + P^Y \frac{\partial^2 Y^e}{\partial M \partial L} \right)}_{b_1} \frac{\partial L}{\partial \omega} + \underbrace{\left( \frac{\partial P^Y}{\partial Y^e} \frac{\partial Y^e}{\partial \omega} \frac{\partial Y^e}{\partial M} + P^Y \frac{\partial^2 Y^e}{\partial M \partial \omega} \right)}_{c_1} =0 \\
    & \underbrace{\left( \frac{\partial P^Y}{\partial Y^e} \frac{\partial Y^e}{\partial M}\frac{\partial Y^e}{\partial L} + P^Y \frac{\partial^2 Y^e}{\partial M \partial L} \right)}_{a_2} \frac{\partial M}{\partial \omega} + \underbrace{\left( \frac{\partial P^Y}{\partial Y^e} \left(\frac{\partial Y^e}{\partial L}\right)^2 + P^Y \frac{\partial^2 Y^e}{\partial L^2} \right)}_{b_2} \frac{\partial L}{\partial \omega} + \underbrace{\left( \frac{\partial P^Y}{\partial Y^e} \frac{\partial Y^e}{\partial \omega} \frac{\partial Y^e}{\partial L} + P^Y \frac{\partial^2 Y^e}{\partial L \partial \omega} \right)}_{c_2} =0, 
\end{align*}
where $b_1=a_2$. Solving the above system of equations, we can arrive at
\begin{equation}\label{eq:dmdomega}
    \frac{\partial M}{\partial \omega} = - \frac{c_1 b_2 - c_2 b_1}{a_1 b_2 - a_2 b_1}.
\end{equation}

We now need to show that the right-hand side of \eqref{eq:dmdomega} above  is non-negative. 
First, consider its denominator. After some algebra, we can show that 
{\small
\begin{align} \label{eq:app_D_a1b2a2b1}
    a_1 b_2 - a_2 b_1 = & P^Y \frac{\partial P^Y}{\partial Y^e} \left[ \frac{\partial^2 Y^e}{\partial M^2} \left(\frac{\partial Y^e}{\partial L}\right)^2 + \left(\frac{\partial Y^e}{\partial M}\right)^2 \frac{\partial^2 Y^e}{\partial L^2} - 2 \frac{\partial Y^e}{\partial M}\frac{\partial Y^e}{\partial L} \frac{\partial^2 Y^e}{\partial M \partial L}  \right] + \left(P^Y\right)^2 \left[ \frac{\partial^2 Y^e}{\partial M^2}\frac{\partial^2 Y^e}{\partial L^2} - \left(\frac{\partial^2 Y^e}{\partial M \partial L}\right)^2 \right].
\end{align}
}This equation has two terms. Under our assumptions of a downward-sloping residual demand function (Assumption \ref{assm:5}) and the production function satisfying the standard neoclassical regularity conditions (Assumption \ref{assm:2}), the first term is non-negative. The second term is a second-order principle minor of the production-function Hessian matrix and, given the concavity of the production function, is non-negative too. Hence, $a_1 b_2 - a_2 b_1 \geq 0$. 

Now, consider the numerator of \eqref{eq:dmdomega}. Noting that $c_1$ and $c_2$ can be rewritten as $P^Y\frac{\partial Y^e}{\partial M} \mu^{-1}$ and $P^Y\frac{\partial Y^e}{\partial L} \mu^{-1}$, respectively, we have 
\begin{align*}
    c_1 b_2 - c_2 b_1 = \mu^{-1} \left(P^Y\right)^2 \left( \frac{\partial Y^e}{\partial M}\frac{\partial^2 Y^e}{\partial L^2} - \frac{\partial^2 Y^e}{\partial M \partial L} \frac{\partial Y^e}{\partial L} \right) \leq 0.
\end{align*}

Therefore, we have shown that $\frac{\partial M}{\partial \omega} \geq 0$.

Similarly, we can show that $\mathcal{M}(\cdot)$ is weakly increasing in $\varphi$. Specifically, by taking partial derivatives of \eqref{eq:foc_prop1} with respect to $\varphi$ and solving for $\frac{\partial M}{\partial \varphi}$, we have
\begin{equation}
    \frac{\partial M}{\partial \varphi} = - \frac{d_1 b_2 - d_2 b_1}{a_1 b_2 - a_2 b_1},
\end{equation}
where $d_1 \equiv \frac{\partial P^Y}{\partial Y^e} \frac{\partial Y^e}{\partial \varphi} \frac{\partial Y^e}{\partial M} + P^Y \frac{\partial^2 Y^e}{\partial M \partial \varphi}$ and $d_2 \equiv \frac{\partial P^Y}{\partial Y^e} \frac{\partial Y^e}{\partial \varphi} \frac{\partial Y^e}{\partial L} + P^Y \frac{\partial^2 Y^e}{\partial L \partial \varphi}$. 

We have already shown that the denominator is non-negative. Then, let us consider the numerator of $\frac{\partial M}{\partial \varphi}$. Recognizing that $\frac{\partial Y^e}{\partial \varphi} = \frac{\partial Y^e}{\partial L} L$, $\frac{\partial^2 Y^e}{\partial M \partial \varphi} = \frac{\partial^2 Y^e}{\partial M \partial L} L$ and $\frac{\partial^2 Y^e}{\partial L \partial \varphi} = \frac{\partial^2 Y^e}{\partial L^2} L + \frac{\partial Y^e}{\partial L}$ and with a few steps of algebra, we have
\begin{equation*}
    d_1 b_2 - d_2 b_1 = - \left( P^Y \right)^2 \frac{\partial Y^e}{\partial L} \left( \frac{1}{\delta} \frac{1}{Y^e} \frac{\partial Y^e}{\partial M}\frac{\partial Y^e}{\partial L} + \frac{\partial^2 Y^e}{\partial M \partial L} \right).
\end{equation*}

It is simple to show that the cross-partial $\frac{\partial^2 Y^e}{\partial M \partial L} = \frac{1}{Y^e} \frac{\partial Y^e}{\partial M}\frac{\partial Y^e}{\partial L} + \frac{Y^e}{LM} \frac{\partial \left(\frac{\partial \ln Y^e}{\partial \ln L} \right)}{\partial \ln M}$. Assuming that $\frac{\partial \left(\frac{\partial \ln Y^e}{\partial \ln L} \right)}{\partial \ln M}=\frac{\partial^2 \ln F}{\partial \ln L\partial \ln M}\ge0$, then $d_1 b_2 - d_2 b_1 = - \left( P^Y \right)^2 \frac{\partial Y^e}{\partial L} \allowbreak \left(\mu^{-1}\frac{1}{Y^e} \frac{\partial Y^e}{\partial M}\frac{\partial Y^e}{\partial L} + \frac{\partial^2 \ln F}{\partial \ln L\partial \ln M} \frac{ Y^e}{LM}\right) \leq 0$, and we have that $\frac{\partial M}{\partial \varphi} \geq 0$.

\vspace{5pt}

\textsl{A non-constant-elasticity demand.}\textemdash In this case, $\mu$ is no longer fixed but a function of $P^Y$. Following the same steps as in the previous case of an isoelastic demand, differentiating the optimality conditions in \eqref{eq:foc_prop1} with respect to $\omega$, we have
\begin{equation}\label{eq:dmdomega2}
    \frac{\partial M}{\partial \omega} = - \frac{c_1' b_2' - c_2' b_1'}{a_1' b_2' - a_2' b_1'},
\end{equation}
where $a_1' = a_1 \mu^{-1} + P^Y \left(\frac{\partial Y^e}{\partial M}\right)^2 \frac{\partial \mu^{-1}}{\partial p^Y} \frac{\partial p^Y}{\partial Y^e}$, 
$b_1' = b_1 \mu^{-1} + P^Y \frac{\partial Y^e}{\partial M} \frac{\partial Y^e}{\partial L} \frac{\partial \mu^{-1}}{\partial p^Y} \frac{\partial p^Y}{\partial Y^e}$,
$c_1' = c_1 \mu^{-1} + P^Y \frac{\partial Y^e}{\partial M} \frac{\partial \mu^{-1}}{\partial p^Y} \frac{\partial p^Y}{\partial Y^e}\frac{\partial Y^e}{\partial \omega}$,
$a_2' = a_2 \mu^{-1} + P^Y \frac{\partial Y^e}{\partial M} \frac{\partial Y^e}{\partial L} \frac{\partial \mu^{-1}}{\partial p^Y} \frac{\partial p^Y}{\partial Y^e}$, 
$b_2' = b_2 \mu^{-1} + P^Y \left(\frac{\partial Y^e}{\partial L}\right)^2 \frac{\partial \mu^{-1}}{\partial p^Y} \frac{\partial p^Y}{\partial Y^e}$,
and $c_2' = c_2 \mu^{-1} + P^Y \frac{\partial Y^e}{\partial L} \frac{\partial \mu^{-1}}{\partial p^Y} \frac{\partial p^Y}{\partial Y^e}\frac{\partial Y^e}{\partial \omega}$. 

First, with some algebra we can show that the denominator in the $\frac{\partial M}{\partial \omega}$ expression in \eqref{eq:dmdomega2} can be written as a sum of two terms:
\begin{equation*}
    a_1' b_2' - a_2' b_1' = (a_1 b_2 - a_2 b_1)\mu^{-2} + \mu^{-1} \left( P^Y \right)^2 \frac{\partial \mu^{-1}}{\partial p^Y} \frac{\partial p^Y}{\partial Y^e} \left[ \frac{\partial^2 Y^e}{\partial M^2} \left(\frac{\partial Y^e}{\partial L}\right)^2 + \frac{\partial^2 Y^e}{\partial L^2} \left(\frac{\partial Y^e}{\partial M}\right)^2 - \allowbreak 2 \frac{\partial^2 Y^e}{\partial M \partial L} \frac{\partial Y^e}{\partial M} \frac{\partial Y^e}{\partial L} \right].
\end{equation*}

Substituting $a_1 b_2 - a_2 b_1$ for \eqref{eq:app_D_a1b2a2b1} and combining the first term of \eqref{eq:app_D_a1b2a2b1} and the second term of $a_1' b_2' - a_2' b_1'$ together, we have
\begin{align*}
    & \left[\mu^{-2} P^Y \frac{\partial P^Y}{\partial Y^e} + \mu^{-1} \left( P^Y \right)^2 \frac{\partial \mu^{-1}}{\partial p^Y} \frac{\partial p^Y}{\partial Y^e} \right] \left[ \frac{\partial^2 Y^e}{\partial M^2} \left(\frac{\partial Y^e}{\partial L}\right)^2 + \frac{\partial^2 Y^e}{\partial L^2} \left(\frac{\partial Y^e}{\partial M}\right)^2 - 2 \frac{\partial^2 Y^e}{\partial M \partial L} \frac{\partial Y^e}{\partial M} \frac{\partial Y^e}{\partial L}\right] \\
    = &  \mu^{-2} \frac{\left(p^Y\right)^2}{Y^e \delta} \left(1 - \frac{\partial \ln \mu}{\partial \ln P^Y} \right) \left[ \frac{\partial^2 Y^e}{\partial M^2} \left(\frac{\partial Y^e}{\partial L}\right)^2 + \frac{\partial^2 Y^e}{\partial L^2} \left(\frac{\partial Y^e}{\partial M}\right)^2 - 2 \frac{\partial^2 Y^e}{\partial M \partial L} \frac{\partial Y^e}{\partial M} \frac{\partial Y^e}{\partial L}\right].
\end{align*}
Given that the second term of \eqref{eq:app_D_a1b2a2b1} is non-negative, if the firm's demand function is such that $\frac{\partial \ln \mu}{\partial \ln P^Y} \leq 1$, we have that $a_1' b_2' - a_2' b_1' \geq 0$.

Next, we sign $c_1' b_2' - c_2' b_1'$, the numerator of \eqref{eq:dmdomega2}. Using the easy-to-establish results that $\frac{\partial Y^e}{\partial \omega} = Y^e$, $\frac{\partial^2 Y^e}{\partial M \partial \omega} = \frac{\partial Y^e}{\partial M}$ and $\frac{\partial^2 Y^e}{\partial L \partial \omega} = \frac{\partial Y^e}{\partial L}$, we can simplify the expressions of $c_1'$ and $c_2'$ as $P^Y \mu^{-1} \frac{\partial Y^e}{\partial M} \left[ 1+ (1+\frac{\partial \ln \mu^{-1}}{\partial \ln P^Y}) \frac{1}{\sigma} \right]$ and $P^Y \mu^{-1} \frac{\partial Y^e}{\partial L} \left[ 1+ (1+\frac{\partial \ln \mu^{-1}}{\partial \ln P^Y}) \frac{1}{\sigma} \right]$, respectively. Also, note that $b_1'$ and $b_2'$ can be rewritten as $b_1' = \mu^{-1}P^Y\times\allowbreak \left( \frac{\partial Y^e}{\partial M} \frac{\partial Y^e}{\partial L} \frac{P^Y}{Y^e} \frac{1}{\sigma} \left( 1 - \frac{\partial \ln \mu}{\partial \ln P^Y} \right) + \frac{\partial^2 Y^e}{\partial M \partial L} \right)$ and $b_2' = \mu^{-1}P^Y\left(  \left(\frac{\partial Y^e}{\partial L}\right)^2 \frac{P^Y}{Y^e} \frac{1}{\sigma} \left( 1 - \frac{\partial \ln \mu}{\partial \ln P^Y} \right) + \frac{\partial^2 Y^e}{\partial L^2} \right)$. With this, we have
\begin{equation*}
  c_1' b_2' - c_2' b_1' = \left( P^Y \mu^{-1} \right)^2 \left[ 1+ \left( 1 - \frac{\partial \ln \mu}{\partial \ln P^Y} \right) \frac{1}{\sigma} \right] \left( \frac{\partial Y^e}{\partial M} \frac{\partial^2 Y^e}{\partial L^2} - \frac{\partial Y^e}{\partial L} \frac{\partial^2 Y^e}{\partial M \partial L} \right).
\end{equation*}
Since $\left( \frac{\partial Y^e}{\partial M} \frac{\partial^2 Y^e}{\partial L^2} - \frac{\partial Y^e}{\partial L} \frac{\partial^2 Y^e}{\partial M \partial L} \right) \leq 0$ as already used earlier, $c_1' b_2' - c_2' b_1' \leq 0$ if the curvature of the demand function is such that $\frac{\partial \ln \mu}{\partial \ln P^Y} \geq 0$.
Putting the numerator and denominator together, we have thus shown that, when $0 \leq \frac{\partial \ln \mu}{\partial \ln P^Y} \leq 1$, we have $\frac{\partial M}{\partial \omega} \geq 0$.

Finally, we show that the conditional material demand is weakly increasing in $\varphi$. To see that, we have
\begin{equation}
    \frac{\partial M}{\partial \varphi} = - \frac{d_1' b_2' - d_2' b_1'}{a_1' b_2' - a_2' b_1'},
\end{equation}
where $d_1' = d_1 \mu^{-1} + P^{Y} \frac{\partial Y^e}{\partial M} \frac{\partial \mu^{-1}}{\partial P^Y} \frac{\partial P^Y}{\partial Y^e} \frac{\partial Y^e}{\partial \varphi}$ and $d_2' = d_2 \mu^{-1} + P^{Y} \frac{\partial Y^e}{\partial L} \frac{\partial \mu^{-1}}{\partial P^Y} \frac{\partial P^Y}{\partial Y^e} \frac{\partial Y^e}{\partial \varphi}$. Through a few steps of simple algebraic manipulation, we can obtain 
\begin{equation*}
    d_1' b_2' - d_2' b_1' = (d_1 b_2 - d_2 b_1) + P^Y \mu^{-2} \frac{\partial \ln \mu}{\partial \ln P^Y} \frac{\partial P^Y}{\partial Y^e}, 
\end{equation*}
which is non-positive if $(d_1 b_2 - d_2 b_1) \leq 0$ and $\frac{\partial \ln \mu}{\partial \ln P^Y} \geq 0$. Therefore, we have shown that $\frac{\partial M}{\partial \varphi} \geq 0$.

This concludes the proof.


\subsection*{Proof of Proposition \ref{prop:2}}

With some necessary adaptations, our proof builds on that of Proposition 3.2 in \citet{demirer2019} for the case with uni-dimensional productivity. 

We first show that, for $(O_{it-1},\widetilde{\mathrm{m}})\in\mathbb{S}$, the likelihood ratio $\frac{f_{\omega}(\omega_{it}|O_{it-1},m_{it}>\widetilde{\mathrm{m}})}{f_{\omega}(\omega_{it}|O_{it-1},m_{it}<\widetilde{\mathrm{m}})}$ satisfies the ``monotone likelihood ratio property,'' i.e.,
\begin{equation*}
\frac{\partial}{\partial \omega_{it}}\left( \frac{f_{\omega}(\omega_{it}|O_{it-1},m_{it}>\widetilde{\mathrm{m}})}{f_{\omega}(\omega_{it}|O_{it-1},m_{it}<\widetilde{\mathrm{m}})}\right)\ge 0.
\end{equation*}
Here, the interest is in $\omega_{it}$ only because $\varphi_{it}$ can be concentrated out of the production function as done in  \eqref{eq:prodfn_tl_novarphi}. To proceed, we rewrite the conditional pdfs inside the ratio using the Bayes rule:
\begin{align}
f_{\omega}(\omega_{it}|O_{it-1},m_{it}>\widetilde{\mathrm{m}}) &= \frac{\Pr(m_{it}>\widetilde{\mathrm{m}}|O_{it-1},\omega_{it})f_{\omega}(\omega_{it}|O_{it-1})}{\Pr(m_{it}>\widetilde{\mathrm{m}}|O_{it-1})}, \\
f_{\omega}(\omega_{it}|O_{it-1},m_{it}<\widetilde{\mathrm{m}}) &= \frac{\Pr(m_{it}<\widetilde{\mathrm{m}}|O_{it-1},\omega_{it})f_{\omega}(\omega_{it}|O_{it-1})}{\Pr(m_{it}<\widetilde{\mathrm{m}}|O_{it-1})}.
\end{align}
Then, the likelihood ratio is given by
\begin{align}
\frac{f_{\omega}(\omega_{it}|O_{it-1},m_{it}>\widetilde{\mathrm{m}})}{f_{\omega}(\omega_{it}|O_{it-1},m_{it}<\widetilde{\mathrm{m}})} &= 
\frac{\Pr(m_{it}>\widetilde{\mathrm{m}}|O_{it-1},\omega_{it})\Pr(m_{it}<\widetilde{\mathrm{m}}|O_{it-1})}{\Pr(m_{it}<\widetilde{\mathrm{m}}|O_{it-1},\omega_{it})\Pr(m_{it}>\widetilde{\mathrm{m}}|O_{it-1})} \notag \\
&= 
\frac{\Pr(m_{it}>\widetilde{\mathrm{m}}|O_{it-1},\omega_{it})\Pr(m_{it}<\widetilde{\mathrm{m}}|O_{it-1})}{(1-\Pr(m_{it}>\widetilde{\mathrm{m}}|O_{it-1},\omega_{it}))\Pr(m_{it}>\widetilde{\mathrm{m}}|O_{it-1})}. \label{eq:likeratio}
\end{align}

The likelihood ratio in \eqref{eq:likeratio} depends on $\omega_{it}$ via probability $\Pr(m_{it}>\widetilde{\mathrm{m}}|O_{it-1},\omega_{it})$ and, clearly, the ratio is increasing in the latter. Therefore, for the likelihood ratio to be weakly increasing in $\omega_{it}$, $\Pr(m_{it}>\widetilde{\mathrm{m}}|O_{it-1},\omega_{it})$ must be weakly increasing in $\omega_{it}$. 

Recall that the material demand function is $M_{it}=\mathcal{M}(K_{it},U_{it},\varphi_{it},\omega_{it})$. Also recognize that, conditional on observable state variables $O_{it-1}$, which include $K_{it-1}$, and the productivities $(\varphi_{it},\omega_{it})'$ that each depend on their lagged values per Markov processes, $K_{it}$ contains no new information because it is predetermined at time $t-1$ and is a deterministic function of $K_{it-1}$ and other past state variables. Abusing notation, we therefore write $M_{it}$ as $M(O_{it-1},U_{it},\varphi_{it},\omega_{it})$ and, correspondingly, $m(O_{it-1},U_{it},\varphi_{it},\omega_{it})$ in logs. Now consider a binary variable $\mathbbm{1}\{m_{it}>\widetilde{\mathrm{m}}\}=\mathbbm{1}\{m(O_{it-1},U_{it},\varphi_{it},\omega_{it})>\widetilde{\mathrm{m}}\}$. We can represent $\Pr(m_{it}>\widetilde{\mathrm{m}}|O_{it-1},\omega_{it})$ as a conditional expectation of this dummy: 
\begin{align}
\Pr(m_{it}>\widetilde{\mathrm{m}}|O_{it-1},\omega_{it}) &= \int\int \mathbbm{1}\{m(O_{it-1},U_{it},\varphi_{it},\omega_{it})>\widetilde{\mathrm{m}}\}f_{U,\varphi}(U_{it},\varphi_{it}|O_{it-1},\omega_{it})dU_{it}d\varphi_{it} \notag \\
&= \int\int \mathbbm{1}\{m(O_{it-1},U_{it},\varphi_{it},\omega_{it})>\widetilde{\mathrm{m}}\}f_{U}(U_{it}|\varphi_{it},O_{it-1},\omega_{it})f_{\varphi}(\varphi_{it}|O_{it-1},\omega_{it})dU_{it}d\varphi_{it} \notag \\
&= \int\int \mathbbm{1}\{m(O_{it-1},U_{it},\varphi_{it},\omega_{it})>\widetilde{\mathrm{m}}\}f_{U}(U_{it}|O_{it-1})f_{\varphi}(\varphi_{it}|O_{it-1},\omega_{it})dU_{it}d\varphi_{it} \notag \\
&= \int \phi(O_{it-1},\varphi_{it},\omega_{it})f_{\varphi}(\varphi_{it}|O_{it-1},\omega_{it})d\varphi_{it},
\end{align}
where we have made use of the joint independence of $U_{it}$ from $(\varphi_{it},\omega_{it})'$ conditional on $O_{it-1}$ per Assumption \ref{assm:5}(ii) in the third line and  have introduced a conditional mean function $\phi(O_{it-1},\varphi_{it},\omega_{it})\equiv  \int \mathbbm{1}\{m(O_{it-1},U_{it},\varphi_{it},\omega_{it})>c\}f_{U}(U_{it}|O_{it-1})dU_{it}$  with the demand heterogeneity $U_{it}$ integrated out in the fourth line.

By Proposition \ref{prop:1}, $m(O_{it-1},U_{it},\varphi_{it},\omega_{it})$ is weakly increasing in both $\omega_{it}$ and $\varphi_{it}$, which implies that $\phi(O_{it-1},\varphi_{it},\omega_{it})$ is also an increasing function of $\omega_{it}$ and  $\varphi_{it}$. Now, take any $\omega_{it}^H\ge \omega_{it}^L$. Owing to the conditional first-order stochastic dominance per Assumption \ref{assm:6}, i.e., $
\mathcal{P}_{\varphi}(\varphi_{it}|\omega_{it}^H,X_{it-1},Z_{it-1})\le\mathcal{P}_{\varphi}(\varphi_{it}|\omega_{it}^L,X_{it-1},Z_{it-1})$, we have that
\begin{align}\label{eq:fosdexp1}
\mathbb{E}\left[\phi\left(O_{it-1},\varphi_{it},\omega_{it}^H\right)|\omega_{it}^H,X_{it-1},Z_{it-1}\right]\ge\mathbb{E}\left[\phi\left(O_{it-1},\varphi_{it},\omega_{it}^H\right)|\omega_{it}^L,X_{it-1},Z_{it-1}\right]
\end{align}
because $\partial\phi(\cdot)/\partial\varphi\ge0$. We also have that
\begin{align}\label{eq:fosdexp2}
\mathbb{E}\left[\phi\left(O_{it-1},\varphi_{it},\omega_{it}^H\right)|\omega_{it}^L,X_{it-1},Z_{it-1}\right]\ge\mathbb{E}\left[\phi\left(O_{it-1},\varphi_{it},\omega_{it}^L\right)|\omega_{it}^L,X_{it-1},Z_{it-1}\right]
\end{align}
because $\partial\phi(\cdot)/\partial\omega\ge0$.

Combining \eqref{eq:fosdexp1} and \eqref{eq:fosdexp2}, we obtain
\begin{align}\label{eq:fosdexp}
\mathbb{E}\left[\phi\left(O_{it-1},\varphi_{it},\omega_{it}^H\right)|\omega_{it}^H,X_{it-1},Z_{it-1}\right]\ge\mathbb{E}\left[\phi\left(O_{it-1},\varphi_{it},\omega_{it}^L\right)|\omega_{it}^L,X_{it-1},Z_{it-1}\right],
\end{align}
and we can then conclude that the mean of $\phi(O_{it-1},\varphi_{it},\omega_{it})$ conditional on $(O_{it-1}',\omega_{it})'$ is weakly increasing in Hicks-neutral productivity $\omega_{it}$ and, therefore, so is  $\Pr(m_{it}>\widetilde{\mathrm{m}}|O_{it-1},\omega_{it})$.

We have thus shown that the likelihood ratio $\frac{f_{\omega}(\omega_{it}|O_{it-1},m_{it}>\widetilde{\mathrm{m}})}{f_{\omega}(\omega_{it}|O_{it-1},m_{it}<\widetilde{\mathrm{m}})}$ is weakly increasing in $\omega_{it}$ thereby satisfying the ``monotone likelihood ratio property.'' In its turn, this property implies the first-order stochastic dominance and the following weak ordering of conditional expectations:
\begin{align}\label{eq:prop32}
\mathbb{E}[\omega_{it}|O_{it-1},m_{it}>\widetilde{\mathrm{m}}] \ge \mathbb{E}[\omega_{it}|O_{it-1},m_{it}<\widetilde{\mathrm{m}}].
\end{align}

Next, substituting for $\omega_{it}$ in \eqref{eq:prop32} using the production function in \eqref{eq:prodfn_tl_novarphi} and recognizing that $\mathbb{E}[\eta_{it}|O_{it-1},m_{it}>\widetilde{\mathrm{m}}]=\mathbb{E}[\eta_{it}|\Xi_{it}]=\mathbb{E}[\eta_{it}]=0$ under Assumption \ref{assm:3}(ii), we obtain
\begin{align}
\mathbb{E}[y_{it}-\overline{y}_{it}(v_{it};\beta)|O_{it-1},m_{it}>\widetilde{\mathrm{m}}] \ge \mathbb{E}[y_{it}-\overline{y}_{it}(v_{it};\beta)|O_{it-1},m_{it}<\widetilde{\mathrm{m}}],
\end{align}
which concludes the proof.


{\small \setlength{\bibsep}{0pt} \bibliography{NonneutralGNRbib} \normalsize}

\begin{thebibliography}{}

\bibitem[Ackerberg et~al., 2020]{afkls2020}
Ackerberg, D., Frazer, G., Kim, K., Luo, Y., and Yingjun, S. (2020).
\newblock Under-identification of structural models based on timing and
  information set assumptions.
\newblock {\em Working Paper}.

\bibitem[Ackerberg et~al., 2007]{abbp2007}
Ackerberg, D.~A., Benkard, C.~L., Berry, S., and Pakes, A. (2007).
\newblock Econometric tools for analyzing market outcomes.
\newblock In Heckman, J.~J. and Leamer, E.~E., editors, {\em Handbook of
  Econometrics}, volume~6A. North Holland.

\bibitem[Ackerberg et~al., 2015]{acf2015}
Ackerberg, D.~A., Caves, K., and Frazer, G. (2015).
\newblock Identification properties of recent production function estimators.
\newblock {\em Econometrica}, 83:2411--–2451.

\bibitem[Baqaee and Farhi, 2019]{BaqaeeFarhi2019}
Baqaee, D.~R. and Farhi, E. (2019).
\newblock {JEEA-FBBVA Lecture 2018: The Microeconomic Foundations of Aggregate
  Production Functions}.
\newblock {\em Journal of the European Economic Association}, 17(5):1337--1392.

\bibitem[Baqaee and Farhi, 2020]{BaqaeeFarhi2020}
Baqaee, D.~R. and Farhi, E. (2020).
\newblock {Productivity and Misallocation in General Equilibrium*}.
\newblock {\em The Quarterly Journal of Economics}, 135(1):105--163.

\bibitem[Battisti et~al., 2022]{battisti2022skill}
Battisti, M., Del~Gatto, M., and Parmeter, C.~F. (2022).
\newblock Skill-biased technical change and labor market inefficiency.
\newblock {\em Journal of Economic Dynamics and Control}, 139:104428.

\bibitem[Brandt et~al., 2017]{brandtetal2017}
Brandt, L., Van~Biesebroeck, J., Wang, L., and Zhang, Y. (2017).
\newblock Wto accession and performance of chinese manufacturing firms.
\newblock {\em American Economic Review}, 107(9):2784--2820.

\bibitem[Canay and Shaikh, 2017]{canayshaikh2017}
Canay, I. and Shaikh, A. (2017).
\newblock Practical and theoretical advances in inference for partially
  identified models.
\newblock In Honor\'e, B., Pakes, A., Piazzesi, M., and Samuelson, L., editors,
  {\em Advances in Economics and Econometrics: Eleventh World Congress}, pages
  271--306. Cambridge University Press.

\bibitem[Chen, 2007]{chen2007}
Chen, X. (2007).
\newblock Large sample sieve estimation of semi-nonparametric models.
\newblock In Heckman, J.~J. and Leamer, E.~E., editors, {\em Handbook of
  Econometrics}, volume~6B. North Holland.

\bibitem[Craven and Wahba, 1979]{cravenwahba1979}
Craven, P. and Wahba, G. (1979).
\newblock Smoothing noisy data with spline functions.
\newblock {\em Numerische Mathematik}, 13:377--403.

\bibitem[{De Loecker}, 2011]{deloecker2011}
{De Loecker}, J. (2011).
\newblock Product differentiation, multiproduct firms, and estimating the
  impact of trade liberalization on productivity.
\newblock {\em Econometrica}, 79:1407--1451.

\bibitem[{De Loecker}, 2013]{deloecker2013}
{De Loecker}, J. (2013).
\newblock Detecting learning by exporting.
\newblock {\em American Economic Journal: Microeconomics}, 5:1--21.

\bibitem[{De Loecker} and Eeckhout, 2020]{deloeckereeckhour2020}
{De Loecker}, J. and Eeckhout, J. (2020).
\newblock Global market power.
\newblock {\em NBER Working Paper No.~ 24768}.

\bibitem[{De Loecker} et~al., 2020]{deloeckeretal2020}
{De Loecker}, J., Eeckhout, J., and Unger, G. (2020).
\newblock The rise of market power and the macroeconomic implications.
\newblock {\em Quarterly Journal of Economics}, 135:561--644.

\bibitem[{De Loecker} et~al., 2016]{deloeckeretal2016}
{De Loecker}, J., Goldberg, P.~K., Khandelwal, A.~K., and Pavcnik, N. (2016).
\newblock Prices, markups, and trade reform.
\newblock {\em Econometrica}, 84:445--510.

\bibitem[{De Loecker} and Warzynski, 2012]{deloeckerwarzynski2012}
{De Loecker}, J. and Warzynski, F. (2012).
\newblock Markups and firm-level export status.
\newblock {\em American Economic Review}, 102:2437--2471.

\bibitem[Demirer, 2019]{demirer2019}
Demirer, M. (2019).
\newblock Production function estimation with imperfect proxies.
\newblock {\em Working Paper, MIT}.

\bibitem[Demirer, 2020]{demirer2020}
Demirer, M. (2020).
\newblock Production function estimation with factor-augmenting technology:
  {A}n application to markups.
\newblock {\em Working Paper, MIT}.

\bibitem[Doraszelski and Jaumandreu, 2013]{dj2013}
Doraszelski, U. and Jaumandreu, J. (2013).
\newblock {R\&D} and productivity: {E}stimating endogenous productivity.
\newblock {\em Review of Economic Studies}, 80:1338--1383.

\bibitem[Doraszelski and Jaumandreu, 2018]{dj2018}
Doraszelski, U. and Jaumandreu, J. (2018).
\newblock Measuring the bias of technological chance.
\newblock {\em Journal of Political Economy}, 126:1027--1084.

\bibitem[Doraszelski and Jaumandreu, 2019]{dj2019}
Doraszelski, U. and Jaumandreu, J. (2019).
\newblock Using cost minimization to estimate markups.
\newblock {\em Working Paper, University of Pennsylvania}.

\bibitem[Flynn et~al., 2019]{flynnetal2019}
Flynn, Z., Gandhi, A., and Traina, J. (2019).
\newblock Measuring markups with production data.
\newblock {\em Working Paper, University of Pennsylvania}.

\bibitem[Gandhi et~al., 2020]{gnr2013}
Gandhi, A., Navarro, S., and Rivers, D. (2020).
\newblock On the identification of gross output production functions.
\newblock {\em Journal of Political Economy}.

\bibitem[Grieco et~al., 2016]{griecoetal2016}
Grieco, P. L.~E., Li, S., and Zhang, H. (2016).
\newblock Production function estimation with unobserved input price
  dispersion.
\newblock {\em International Economic Review}, 57:665--689.

\bibitem[Grieco et~al., 2020]{griecopaint}
Grieco, P. L.~E., Li, S., and Zhang, H. (2020).
\newblock Input prices, productivity and trade dynamics: {L}ong-run effects of
  liberalization on {C}hinese paint manufacturers.
\newblock {\em Working Paper}.

\bibitem[Griliches and Mairesse, 1998]{gm1998}
Griliches, Z. and Mairesse, J. (1998).
\newblock Production functions: {T}he search for identification.
\newblock In {\em Econometrics and Economic Theory in the Twentieth Century:
  The Ragnar Frisch Centennial Symposium}, pages 169--203. Cambridge University
  Press.

\bibitem[Kim et~al., 2019]{kimetal2019}
Kim, K., Luo, Y., and Su, Y. (2019).
\newblock A robust approach to estimating production functions: {R}eplication
  of the {ACF} procedure.
\newblock {\em Journal of Applied Econometrics}, 34:612--619.

\bibitem[Kline et~al., 2021]{klineetal2021}
Kline, B., Pakes, A., and Tamer, E. (2021).
\newblock Moment inequalities and partial identification in industrial
  organization.
\newblock {\em NBER Working Paper \#29409}.

\bibitem[Levinsohn and Petrin, 2003]{lp2003}
Levinsohn, J. and Petrin, A. (2003).
\newblock Estimating production functions using inputs to control for
  unobservables.
\newblock {\em Review of Economic Studies}, 70:317--341.

\bibitem[Malikov and Lien, 2021]{malikovlien}
Malikov, E. and Lien, G. (2021).
\newblock Proxy variable estimation of multiproduct production functions.
\newblock {\em American Journal of Agricultural Economics}, 103(5):1878--1902.

\bibitem[Malikov et~al., 2021]{malikov2021jom}
Malikov, E., Zhang, J., Zhao, S., and Kumbhakar, S.~C. (2021).
\newblock Accounting for cross-location technological heterogeneity in the
  measurement of operations efficiency and productivity.
\newblock {\em Journal of Operations Management}.
\newblock forthcoming.

\bibitem[Malikov and Zhao, 2021]{malikovzhao2021}
Malikov, E. and Zhao, S. (2021).
\newblock On the estimation of cross-firm productivity spillovers with an
  application to {FDI}.
\newblock {\em Review of Economics and Statistics}.
\newblock forthcoming.

\bibitem[Malikov et~al., 2020]{malikovetal2020}
Malikov, E., Zhao, S., and Kumbhakar, S.~C. (2020).
\newblock Estimation of firm-level productivity in the presence of exports:
  {E}vidence from {C}hina's manufacturing.
\newblock {\em Journal of Applied Econometrics}, 35:457--480.

\bibitem[Mammen, 1993]{mammen1993}
Mammen, E. (1993).
\newblock Bootstrap and wild bootstrap for high dimensional linear models.
\newblock {\em Annals of Statistics}, 21:255--285.

\bibitem[Mo et~al., 2021]{moetal2021}
Mo, J., Qiu, L.~D., Zhang, H., and Dong, X. (2021).
\newblock What you import matters for productivity growth: {E}xperience from
  {C}hinese manufacturing firms.
\newblock {\em Journal of Development Economics}, 152.
\newblock Article 102677.

\bibitem[Molinari, 2020]{molinari2020}
Molinari, F. (2020).
\newblock Microeconometrics with partial identification.
\newblock {\em Working Paper CWP15/20, CEMMAP, The Institute for Fiscal
  Studies, Department of Economics, UCL}.

\bibitem[Newey, 1984]{newey1984}
Newey, W.~K. (1984).
\newblock A method of moments interpretation of sequential estimators.
\newblock {\em Economics Letters}, 14(2):201--206.

\bibitem[Oberfield and Raval, 2021]{oberfieldravel2021}
Oberfield, E. and Raval, D. (2021).
\newblock Micro data and macro technology.
\newblock {\em Econometrica}, 89(2):703--732.

\bibitem[Olley and Pakes, 1996]{op1996}
Olley, G.~S. and Pakes, A. (1996).
\newblock The dynamics of productivity in the telecommunications equipment
  industry.
\newblock {\em Econometrica}, 64:1263--1297.

\bibitem[Raval, 2020]{raval2020}
Raval, D. (2020).
\newblock Testing the production approach to markup estimation.
\newblock {\em Working Paper, Federal Trade Commission}.

\bibitem[Rothenberg, 1971]{rothenberg1971}
Rothenberg, T.~J. (1971).
\newblock Identification in parametric models.
\newblock {\em Econometrica}, 39:577--591.

\bibitem[Stoye, 2021]{stoye2021}
Stoye, J. (2021).
\newblock A simple, shot, nut never-empty confidence interval for partially
  identified parameters.
\newblock {\em Working Paper, Cornell University}.

\bibitem[{Van Biesebroeck}, 2005]{vanbiesebroeck2005}
{Van Biesebroeck}, J. (2005).
\newblock Exporting raises productivity in sub-{S}aharan {A}frican
  manufacturing firms.
\newblock {\em Journal of International Economics}, 67:373--391.

\bibitem[Wooldridge, 2009]{wooldridge2009}
Wooldridge, J.~M. (2009).
\newblock On estimating firm-level production functions using proxy variables
  to control for unobservables.
\newblock {\em Economics Letters}, 104:112--114.

\bibitem[Zhang, 2017]{zhang2017eer}
Zhang, H. (2017).
\newblock Static and dynamic gains from costly importing of intermediate
  inputs: {Evidence from Colombia}.
\newblock {\em European Economic Review}, 91:118--145.

\bibitem[Zhang, 2019]{zhang2019jde}
Zhang, H. (2019).
\newblock Non-neutral technology, firm heterogeneity, and labor demand.
\newblock {\em Journal of Development Economics}, 140:145--168.

\bibitem[Zhao et~al., 2020]{zhaoetal2020}
Zhao, S., Qian, B., and Kumbhakar, S.~C. (2020).
\newblock Estimation of productivity and markups with price dispersion:
  Evidence from chinese manufacturing during economic transition.
\newblock {\em Southern Economic Journal}, 87(2):666--699.

\end{thebibliography}


\end{document}